\documentclass[12pt]{iopart}
\usepackage{graphicx,amssymb,iopams}
\expandafter\let\csname equation*\endcsname\relax
\expandafter\let\csname endequation*\endcsname\relax
\usepackage{amsmath,caption,array,float,subfigure,amsmath,amsfonts}
\usepackage{subfigure,epstopdf,epsfig,graphicx,caption}
\usepackage{comment,xcolor}
\newtheorem{theorem}{Theorem}[section]
\newtheorem{lemma}[theorem]{Lemma}
\newtheorem{example}{Example}[section]
\newtheorem{proof}{Proof}[section]
\begin{document}


\title[]{Parametric excitation and Hopf bifurcation analysis of a time delayed nonlinear feedback oscillator}

\author{Sandip Saha$^1$, Gautam Gangopadhyay$^1$, Sangeeta Kumari$^2$ and Ranjit Kumar Upadhyay$^2$}
\address{$^1$S N Bose National Centre For Basic Sciences\\ ~~Block-JD, Sector-III, Salt Lake, Kolkata-700106, India\\~\\
$^2$Department of Mathematics \& Computing,\\ ~~Indian Institute of Technology (Indian School Mines),\\ ~~Dhanbad-826004, India}


\date{\today}

\begin{abstract}
In this paper, an attempt has been made to understand the parametric excitation of a periodic orbit of nonlinear oscillator which can be a limit cycle, center or a slowly decaying center-type oscillation. For this a delay model is considered with nonlinear feedback oscillator defined in terms of Li\'enard oscillator description which can give rise to any one of the periodic orbits stated above. We have characterized the resonance and antiresonance behaviour for arbitrary nonlinear system  from their stability and bifurcation analyses in reference to  the standard delayed van der Pol system. An approximate analytical solution using Krylov--Bogoliubov (K-B) averaging method is utilised to recognize the sub-harmonic resonance and antiresonance, and average energy consumption per cycle. Direction of Hopf bifurcation and stability of the periodic solution bifurcating from the trivial fixed point are carried out using normal form and center manifold theory. The parametric excitation is also thoroughly investigated via bifurcation analysis to find the role of the control parameters like time delay, damping and nonlinear terms.
\end{abstract}

\textbf{Keywords:} Parametric resonance, Li\'enard oscillator, Limit cycle, Krylov--Bogoliubov averaging method, Time delay, Bifurcation analysis

\section{Introduction}  \label{sec1}
Dynamical systems~\cite{strogatz,nayfeh,holmes,birkhoff,len0} capable of having isochronous~\cite{calogero,Sarkariso2012} oscillations are very important from the point of view of modelling real world systems which exhibit self-sustained oscillations~\cite{Jenkins_2013}. Isochronicity~\cite{calogero} is a widely studied subject not only for its relation with stability theory and bifurcation theory~\cite{strogatz,holmes,rkupadhyay} but also in the discrimination of oscillating orbits in limit cycle, center and center-type focus~\cite{powerlaw,len3.5} in terms of Li\'enard equation~\cite{strogatz,remickens,len4,Saha2019}. Many nonlinear dynamical systems in various scientific disciplines are influenced by the finite propagation time of signals in feedback loops modelled with time-delay~\cite{goto2007renormalization,atay}. In some systems, such as lasers and electro mechanical models, large varieties of delay appear~\cite{algaba,hu,jiang,wang,xu} with  van der Pol oscillator~\cite{strogatz,algaba,limiso} is a standard prototype with delayed feedback have very rich and complex bifurcation~\cite{jiang,wang,xu}. A classical van der Pol equation with delayed feedback has been extensively investigated having Bogdanov--Taken bifurcation with triple zero and Hopf-zero singularity~\cite{jiang} as well as transcritical and pitchfork bifurcations~\cite{wang,Cooke_Grossman_82}. In a similar context stability analysis of a pair of van der Pol oscillators with delayed self-connection, position and velocity couplings are also performed~\cite{hu}.

{\color{black}The existence of a limit cycle can be predicted by the Poincar\'e--Bendixson theorem, but the exact location and size of the limit cycle cannot be predicted in advance~\cite{strogatz}. This is why perturbation theory is employed to know the finer details about the shape of the limit cycles for various types of Li\'enard oscillators although the procedure of approaching the problem perturbatively becomes non-trivial. A tremendous progress has been made in the area of dissipative dynamical systems regarding the applicability of the multiscale perturbative methods~\cite{strogatz,len0,len3.5,remickens,chen1,chen2,len3,Wilson1976,slross} like Krylov--Bogoliubov (K-B), Lindstedt--Poincar\'e and Renormalisation Group (RG). But, except the RG approach other methods are mostly restricted to weak nonlinearity~\cite{strogatz,nayfeh,len0,slross}. In a recent development  Sarkar et al.~\cite{len3.5} have attempted to understand the application of the RG principle to probe the difference between limit cycle and center through problems in dynamics for various 2-D systems.

One can also characterize limit cycle through the parametric excitation~\cite{penvo,momeni} as this type of excited system is prepared by time-varying coefficients of the equations of motion which plays a drastically different role compared to the usual direct external driving force through an additive term. Recently, for a van der Pol oscillator the effect of periodically modulating nonlinear term has been investigated as a parametrically excited nonlinearity which can provide the phenomenon of both resonance and antiresonance depending on the frequency of the parametrical drive. Again, time delay even in a simple nonlinear system as a feedback~\cite{Balanov2005} device induces a large variety of dynamical scenario, like tori and new chaotic attractors along with the fact that the delay modifies the periods and the stabilities of the limit cycles in the system depending on the strength of the feedback and its magnitude. Stability of a periodic orbit, namely a limit cycle or a center using a controlled delay is quite interesting and that too in presence of a parametric excitation leading to resonance and antiresonance is very useful for arbitrary quartic nonlinear system as a generalization of van der Pol system~\cite{penvo} with a limit cycle.}

Inspite of a great deal of investigations on parametric driving, the response function due to a parametric excitation of an arbitrary  periodic orbit is not well understood. Here we have investigated a delayed nonlinear system  to obtain both limit cycle, center and a slowly decaying  center-type~\cite{powerlaw,len3.5} oscillation for different range of parameters of the proposed model. The model exhibits a very rich dynamics due to the presence of delay which was first introduced by Goto~\cite{goto2007renormalization} to apply renormalization method~\cite{len3.5,goto2007renormalization,limiso,chen1,chen2,len3} to compare between perturbative analysis with the exact solution and later Sarkar et al.~\cite{len3.5,len3} used it in a similar context. The extended version of the model is analysed by Saha et al.~\cite{powerlaw} through K-B approach exhibiting  a stable limit cycle due to the presence of weak delay. As a further extension with a delayed feedback which becomes a variant of delayed van der Pol system, we have given here an approximate solution from the multiscaled perturbation theory which  is utilised to characterise the  parametric resonance and antiresonance behaviour. This is also more elaborately dealt by performing the bifurcation analysis~\cite{holmes,rkupadhyay,Cooke_Grossman_82,hassard} of the system due to parametric excitation. As there is no general scheme to handle a delayed system using perturbation theory~\cite{nayfeh,len0} to study bifurcation or periodic orbit and characterisation of the properties of nonlinear oscillation, in this paper we have analysed the Hopf bifurcation scenario during the parametric excitation  to probe the dynamics due to the interplay of nonlinear damping and delay terms.

The layout of the paper is as follows. In Section \ref{sec2}, the nonlinear delay model is introduced along with the K-B analysis. Parametric excitation is studied in Section \ref{sec3}. In Section \ref{sec4}, a detailed stability and bifurcation analysis including direction and stability of Hopf bifurcation have been performed. Numerical simulations are executed in Section \ref{sec5}. Finally, the paper is concluded in Section \ref{sec7}.

\section{Time Delayed  Nonlinear Feedback Oscillator} \label{sec2}
In this section, we consider a model of delay dynamics  where the oscillation is fed energy through delay term in potential and its total energy increases with time. To control the feedback effect   we have introduced a linear damping term $b$, which gives a center solution similar to simple harmonic oscillator and then in presence of damping and delay we have introduced another quartic nonlinear term $a$. By this way  we have introduced a variant of delayed van der Pol system  which gives both limit cycle and center-type solution by tuning the parameters. The basic equations of the model are
\begin{align}
  \dot{x}(t) &= y(t), \nonumber\\
  \dot{y}(t) &= - \epsilon \lbrace x(t-t_d)+(a x^2(t)+b) \dot{x}(t) \rbrace-\omega^2 x(t).      
  \label{eq}
\end{align}
Here we have considered, $0<\epsilon \ll 1$ is a small perturbative constant and ($0,0$) is the only fixed point. From the above model we can find the delayed  van der Pol case of a limit cycle if we consider $b=-1$ and the sign of the term $\epsilon x(t-t_d)$ is negative. In absence of $a$ and $b$, the system converts into a  delayed harmonic oscillator which was introduced by Goto~\cite{goto2007renormalization} to apply conventional renormalization method to show an agreement between  perturbative analysis with the exact solution where the oscillation is fed energy through delay term and its total energy increases with time. Further, in absence of $a$ i.e. $a=0$, a detailed bifurcation analysis in presence of delay for a linear-damped-delayed system has provided by Cooke et al.~\cite{Cooke_Grossman_82}(Eq.~(10), pp-602) through stability analysis. The extended version i.e. Eq.~\eqref{eq} has been analysed by Saha et al.~\cite{powerlaw} through K-B approach which exhibits a stable limit cycle, a center or slowly decaying center-type dynamics due to the presence of weak delay.

In what follows, the challenges of the Hopf bifurcation analysis for such kind of weak delayed system are investigated in details and we have provided a straight forward way to bifurcation analysis through amplitude equation using K-B averaging, which is much easier than traditional Hopf bifurcation analysis. 

\subsection{Approach through Krylov--Bogoliubov (K-B) Averaging}
\label{subsection 2}
This time delayed system can be written as,
\begin{align}
\ddot{x}(t)+\epsilon h+\omega^2 x(t) =0;~~~ h=x(t-t_d)+(a x^2+b) \dot{x}(t).\label{eq4}
\end{align}

According to K-B scheme for approximate analytical solution of an ordinary differential equation, we consider, $x(t)=r(t) \hspace{0.1 cm} \cos (\omega t+\phi(t))$  with $r(t) = \sqrt{x^2 +\frac{\dot{x}^2}{\omega^2}}$ and $\phi(t) = -\omega t+ \tan^{-1} (- \frac{\dot{x}}{\omega x})$. Then one can obtain $\dot{r}(t) = \frac{\epsilon h}{\omega} \sin (\omega t+\phi(t))$ and $\dot{\phi}(t) = \frac{\epsilon h}{\omega r(t)} \cos (\omega t+\phi(t))$ i.e., the time derivative of  amplitude and phase are of $O(\epsilon)$.

Taking running average of $\phi$ dependent functions, $U$ defined as,
\begin{align}
\overline{U}(t) = \frac{\omega}{2 \pi} \int_{t-\frac{\pi}{\omega}}^{t+\frac{\pi}{\omega}} U(s) ds,  \label{3}
\end{align} 
where $\omega$ is the natural frequency of the system and considering $\dot{\overline{U}}=\overline{\dot{U}}$ from the fundamental theorem of calculus,
 one can obtain,
\begin{align}
\dot{\overline{r}} &= \left\langle \frac{\epsilon}{\omega} \hspace{.15cm} h(x,y)\hspace{.15cm} \sin (\omega t+\phi(t)) \right\rangle_t,		\nonumber  \\
\dot{\overline{\phi}} &= \left\langle \frac{\epsilon \hspace{.15cm} h(x,y)\hspace{.15cm}}{\omega r(t)} \cos (\omega t+\phi(t)) \right\rangle_t. \label{eq7}
\end{align}

Since $\dot{r}(t)$ and $\dot{\phi}(t)$ are of $O(\epsilon)$ then we can set the perturbation on $r$ and $\phi$ over one cycle as,
\begin{align}
r(t) &=\overline{r}+O(\epsilon),	\nonumber\\
\phi(t) &=\overline{\phi}+O(\epsilon),	\label{eq8}
\end{align}
where $\overline{r}$ and $\overline{\phi}$ are not exactly constant, they are very weakly $t$-dependent so that the error can be negligible.

Now using all the approximations, $h$ reduces to $h= \overline{r}  \lbrace \cos(\omega t+\overline{\phi})\cos(\omega t_d)+\sin(\omega t+\overline{\phi})\sin(\omega t_d)-\omega (a \overline{r}^2 \cos^2(\omega t+\overline{\phi})+b)\sin(\omega t+\overline{\phi}) \rbrace+O(\epsilon)$, where one finds $r(t-t_d)$ is approximated as $r(t-t_d)=r(t)-t_d \dot{r}(t)=r(t)+O(\epsilon)$ as $\dot{r}(t)=O(\epsilon)$ and $0<t_d \ll1$. Finally one can obtain,
\begin{align}
\dot{\overline{r}} &= -\frac{\epsilon \overline{r}}{8} \left \lbrace a \overline{r}^2-4\left(\frac{\sin(\omega t_d)}{\omega}-b\right)\right \rbrace, \nonumber\\
\dot{\overline{\phi}} &= \frac{\epsilon}{2 \omega} \cos(\omega t_d),	\label{eq10}
\end{align}
where $O(\epsilon^2)$ terms are neglected. Now, if we have $(x_0, y_0)$ as an initial condition of the system and then finding, $r_0$ and $\phi_0$, and one can find the approximate solution by solving $\dot{\overline{r}}$ and $\dot{\overline{\phi}}$. Also, one finds the energy as $E=\frac{\omega^2 x^2+\dot{x}^2}{2}$ to get  the energy consumption over one cycle, which is $\Delta E=2 \pi \overline{r}\hspace{0.05cm} \dot{\overline{r}}$.
 
Now to focus on the bifurcation point for the above type of delayed feedback oscillator, bifurcation analysis is not meaningful through linear perturbation in $\epsilon$ around the fixed point ($0,0$), as it truncates the delay. For example, if we put $x=0+\epsilon X$ and $\dot{x}=y=0+\epsilon Y$ to \eqref{eq} and compare to the $O(\epsilon)$ to figure out the Jacobian matrix as well as the characteristic equation then it does not include the effect of delay in the characteristic equation. In this situation we get a center solution, where the physical system is something different from the original one. Now coming to the point, which is, the simple delayed feedback case with ($a=0,b=0$), bifurcation is far away from it. Again, when $a=0$, one can find the increasing (when system diverges) or decreasing energy (when system undergoes a focus with a decaying center) according to $b<\frac{\sin(\omega t_d)}{\omega}$ and $b>\frac{\sin(\omega t_d)}{\omega}$, respectively. This provides that the value of $b=\frac{\sin(\omega t_d)}{\omega}$ is the bifurcating point having the center solution. Similarly, for the case of non-zero value of $a$ i.e., in presence of nonlinearity, if $b<\frac{\sin(\omega t_d)}{\omega}$ then the system has a non-zero radius providing the stable limit cycle solution and if $b>\frac{\sin(\omega t_d)}{\omega}$ having imaginary radius giving the unstable limit cycle (asymptotically stable fixed point). So,  $b=\frac{\sin(\omega t_d)}{\omega}$ is the Hopf bifurcating parameter, defining the energy transfer zone between stable and unstable focus connecting a center and slowly decaying center-type solution. Numerical simulation is given below for better understanding.

\begin{figure}[!ht]
\begin{center}
\includegraphics*[width=0.75\linewidth]{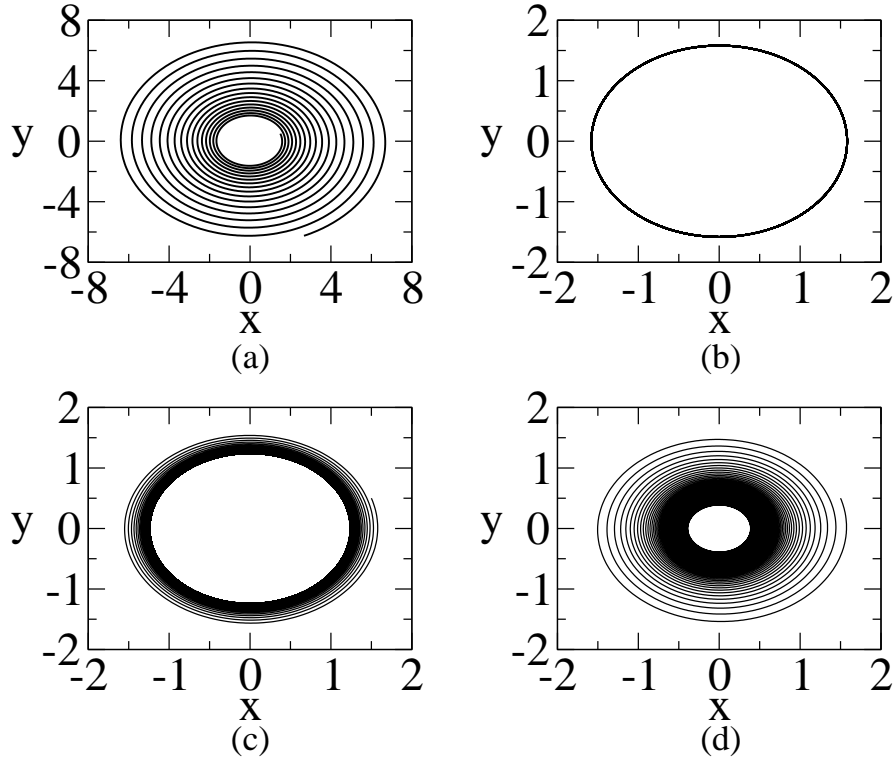}
\caption{\textbf{Time-delayed system:} Phase space plots of Eq.~\eqref{eq} are shown using approximate amplitude and phase equations with time delay, $t_d=0.623$ where (a) $a=0,b=0$ for a feedback system with increasing phase space area, (b) center with $a=0,b=\frac{\sin(\omega t_d)}{\omega}$, (c) LC with $a=1,b=0.2<\frac{\sin(\omega t_d)}{\omega}$ and (d) center-type orbit with $a=1,b=\frac{\sin(\omega t_d)}{\omega}$ with $\epsilon=0.05$ and $\omega=1$.} \label{fig1}
\end{center}
\end{figure}

Fig.~\ref{fig1} shows the corresponding numerical simulations of Eq.~\eqref{eq} owing to four situations, where (a) is the case for the delay with no nonlinearity and damping, i.e, $a=0,b=0$, which shows a simple feedback oscillator with continuously increasing energy in the system. Further, (b), (c) and (d) shows a center with $a=0,b=\frac{\sin(\omega t_d)}{\omega}$, a limit cycle(LC) with $a=1,b=0.2<\frac{\sin(\omega t_d)}{\omega}$ and a slowly decaying center-type orbit or focus with $a=1,b=\frac{\sin(\omega t_d)}{\omega}$, respectively. Here the finite delay, $t_d$ and $\epsilon$ are kept fixed at $0.623$ and $0.05$, respectively along with unit frequency.
 
\section{Parametrically Excited Time Delayed Nonlinear Feedback Oscillator}  \label{sec3}
As an application of the approximate analytical solution and bifurcation situation, in this section, we have investigated the time delayed nonlinear oscillator under parametric excitation. The resonance and antiresonance behaviours for the limit cycle, center and center-type orbits are described both analytically and numerically for weak delayed situation. The stability regions at the resonances of the system under excitation are explored.

\subsection{Periodic Solutions for Resonance and Antiresonance Cases}	\label{subsec1}
Making the system parametrically excited~\cite{penvo,momeni} by  a periodic force $\cos(\Omega t)$ with a weighted constant, $\gamma$, the basic equations become:
\begin{align}
\dot{x}(t) =& y(t), 	\nonumber\\
\dot{y}(t) =& - \epsilon \lbrace x(t-t_d)+[1+\gamma \cos(\Omega t)] (a x^2(t)+b) \dot{x}(t) \rbrace-\omega^2 x(t); \quad \gamma \neq 0,  \Omega \in \mathbb{Z}_{\neq 0},  \label{7}
\end{align}
where $0 < \epsilon \ll 1$, $a,b$ are the system parameters and $t_d(0<t_d \ll 1)$ is indicating the time delay.

To have an approximate analytical solution of the above system \eqref{7} one can use the K-B averaging method which can be the asymptotically stable, limit cycle, center and center-type solution depending upon the parameters. But, in this case calculations will be quite harder due to the non zero excitation strength and one can not get it naturally like the previous one due to some singularities as well as heavily coupled nonlinear functional forms of the amplitude and phase equations. K-B averaging gives,
\begin{align}
\dot{\overline{r}} =& -\frac{\epsilon \overline{r}}{8} \left\lbrace a \overline{r}^2-4\left(\frac{\sin (\omega t_d)}{\omega}-b\right) \right\rbrace +A(\overline{r},\overline{\phi}), 	\nonumber \\
\dot{\overline{\phi}} =& \frac{\epsilon}{2 \omega} \cos(\omega t_d)+B(\overline{r},\overline{\phi}),	\label{8}
\end{align}
where, $A(\overline{r},\overline{\phi})=A_1(\overline{r},\overline{\phi})+A_2(\overline{r},\overline{\phi})$ and $B(\overline{r},\overline{\phi})=B_1(\overline{r},\overline{\phi})+B_2(\overline{r},\overline{\phi})$ with

{\scriptsize
\begin{align*}
A_1(\overline{r},\overline{\phi}) =& \frac{b \gamma  \overline{r} \omega  \epsilon}{2 \pi  \left(8 \omega ^2 \Omega -2 \Omega ^3\right)}  \left(\sin \left(\frac{2 \pi  \Omega }{\omega }\right) \left(-4 \omega ^2+\Omega ^2-\Omega ^2 \cos (2 \overline{\phi} )\right)-8 \omega  \Omega  \sin (\overline{\phi} ) \cos (\overline{\phi} ) \sin ^2\left(\frac{\pi  \Omega }{\omega }\right)\right),\\
A_2(\overline{r},\overline{\overline{\phi}})= &-\frac{a \gamma  \overline{r}^3 \omega  \epsilon}{16 \pi  \left(\Omega ^3-16 \omega ^2 \Omega \right)}  \left(\sin \left(\frac{2 \pi  \Omega }{\omega }\right) \left(-16 \omega ^2+\Omega ^2-\Omega ^2 \cos (4 \overline{\phi} )\right)-8 \omega  \Omega  \sin (4 \overline{\phi} ) \sin ^2\left(\frac{\pi  \Omega }{\omega }\right)\right);\\
B_1(\overline{r},\overline{\phi}) =& \frac{a \gamma  \overline{r}^2 \omega  \epsilon}{32 \pi }  \left(-\frac{8 \omega  \cos (2 \overline{\phi} )}{4 \omega ^2-\Omega ^2}-\frac{8 \omega  \cos (4 \overline{\phi} )}{16 \omega ^2-\Omega ^2}+\frac{2 \cos \left(2 \left(\frac{\pi  \Omega }{\omega }+\overline{\phi} \right)\right)}{2 \omega +\Omega }+\frac{\cos \left(\frac{2 \pi  \Omega }{\omega }+4 \overline{\phi} \right)}{4 \omega +\Omega }+\frac{2 \cos \left(2 \overline{\phi} -\frac{2 \pi  \Omega }{\omega }\right)}{2 \omega -\Omega }+\frac{\cos \left(4 \overline{\phi} -\frac{2 \pi  \Omega }{\omega }\right)}{4 \omega -\Omega }\right),\\
B_2(\overline{r},\overline{\phi}) =& \frac{b \gamma  \omega  \epsilon}{2 \pi  \left(8 \omega ^2-2 \Omega ^2\right)}  \left(\Omega  \sin (2 \overline{\phi} ) \sin \left(\frac{2 \pi  \Omega }{\omega }\right)-4 \omega  \cos (2 \overline{\phi} ) \sin ^2\left(\frac{\pi  \Omega }{\omega }\right)\right).
\end{align*}}

There are two singularities in the equation of $\dot{\overline{r}}$ and $\dot{\overline{\phi}}$ which are at $\Omega = 2\omega, 4\omega$. One can take limit for amplitude and phase equation near the singularities and get two different sets of equations, are
\begin{align}
\dot{\overline{r}}|_{\Omega \rightarrow 2\omega} =& -\frac{\epsilon \overline{r}}{8} \left \lbrace a \overline{r}^2-4\left(\frac{\sin (\omega t_d)}{\omega}-b\right)\right\rbrace  +\frac{1}{4} b \gamma  \overline{r} \epsilon  \cos (2 \overline{\phi} ),	\nonumber\\
\dot{\overline{\phi}}|_{\Omega \rightarrow 2\omega} =& \frac{\epsilon}{2 \omega} \cos(\omega t_d)-\frac{1}{8} \gamma  \epsilon  \sin (2 \overline{\phi} ) \left(a \overline{r}^2+2 b\right),		\label{9}
\end{align}
and
\begin{align}
\dot{\overline{r}}|_{\Omega \rightarrow 4\omega} =& -\frac{\epsilon \overline{r}}{8} \left\lbrace a \overline{r}^2-4\left(\frac{\sin (\omega t_d)}{\omega}-b\right)\right\rbrace  +\frac{1}{16} a \gamma  \overline{r}^3 \epsilon  \cos (4 \overline{\phi} ),		\nonumber\\
\dot{\overline{\phi}}|_{\Omega \rightarrow 4\omega} =& \frac{\epsilon}{2 \omega} \cos(\omega t_d)-\frac{1}{16} a \gamma  \overline{r}^2 \epsilon  \sin (4 \overline{\phi} ),		\label{10}
\end{align}
where the non-zero $\gamma$ carries the additional part. Note that, here the systems are coupled and autonomous. They can be solved analytically by taking some additional conditions in the phase lag or by numerically in a synchronized state. Also, one can get all likely features obtained in the previous section but the distinction of limit cycle, center and center-type situations due to the tuning of system parameters are harder little a bit, which is discussed in the numerical simulation section.

To have an idea about the excitation strength $\gamma$ we are now transforming the polar equations into a van der Pol plane for each modes of resonances. It is pre-assumed that the unique trivial fixed point will remain unstable or neutral as we are focussing upon the limit cycle or center or center-type situations. Also, another reason to consider a van der Pol plane is that one can not perform the traditional linear stability analysis for the excited Li\'enard system~\eqref{7}. So to perform the same taking a trial solution, $x(t)=u(t) \cos(\omega t)+v(t) \sin(\omega t)$ and the one can obtain two simplified sets of equations as:

\begin{align}
\dot{u}|_{\Omega\rightarrow2\omega}&= \frac{\epsilon}{8 \omega }  \left(u \left(-a \omega  \left(u^2+(1-2 \gamma ) v^2\right)+2 b (\gamma -2) \omega +4 \sin \left(\omega  t_d\right)\right)+4 v \cos \left(\omega  t_d\right)\right), \nonumber \\
\dot{v}|_{\Omega\rightarrow2\omega} &= -\frac{\epsilon}{8 \omega }  \left(v \left(a v^2 \omega +2 b (\gamma +2) \omega -4 \sin \left(\omega  t_d\right)\right)+a (2 \gamma +1) u^2 v \omega +4 u \cos \left(\omega  t_d\right)\right),		\label{eqo2}
\end{align}
and
\begin{align}
\dot{u}|_{\Omega\rightarrow4\omega}&=  \frac{\epsilon}{16 \omega }  \left(u \left(8 \left(\sin \left(\omega  t_d\right)-b \omega \right)-a (3 \gamma +2) v^2 \omega \right)+a (\gamma -2) u^3 \omega +8 v \cos \left(\omega  t_d\right)\right),	\nonumber \\
\dot{v}|_{\Omega\rightarrow4\omega}&= \frac{\epsilon}{16 \omega }  \left(-a (3 \gamma +2) u^2 v \omega +a (\gamma -2) v^3 \omega +8 v \left(\sin \left(\omega  t_d\right)-b \omega \right)-8 u \cos \left(\omega  t_d \right)\right).	\label{eqo4}
\end{align}
 
Here we have approximated $u(t-t_d)=u(t)-t_d \dot{u}(t)$ and $v(t-t_d)=v(t)-t_d \dot{v}(t)$ where $O(t_d^2)$, $\ddot{u}(t), \ddot{v}(t),\epsilon \dot{u}(t), \epsilon\dot{v}(t)$ are neglected. Note that, the above sets of equations are the first order autonomous equations and are symmetrical or even in nature with  $(u, v) \rightarrow (-u,-v)$. Now the traditional stability analysis of Eqs.~\eqref{eqo2} and \eqref{eqo4} can be done by linearising around the fixed point $(0,0)$ which is the common fixed point of both the Eqs.~\eqref{eqo2} and \eqref{eqo4} as well as the original Eq.~\eqref{eq}. It is easily deduced from Eqs.~\eqref{eqo2} and \eqref{eqo4} that $u = v = 0$ (i.e., $x = 0$) is an equilibrium solution determines the fixed point. The linear stability analysis of Eqs.~\eqref{eqo2} and \eqref{eqo4} will help us to find the valid excitation strength under the corresponding state of limit cycle, center  or center-type cases.

\subsection{Numerical Results for Parametric Excitation}

Here we have numerically characterized the resonance and antiresonance behaviours for the limit cycle, center and center-type cases. Plots of amplitude variations are provided with $\gamma$ for two different resonances at $\Omega=2$(black, doted) and $\Omega=4$(red) in terms of the scaled radius of the orbits. Further, phase space plots, amplitude variations with time and average energy consumption per cycle at their steady state behaviours of each states are also shown. 

\begin{figure}[h]
\centering
\includegraphics*[width=\linewidth]{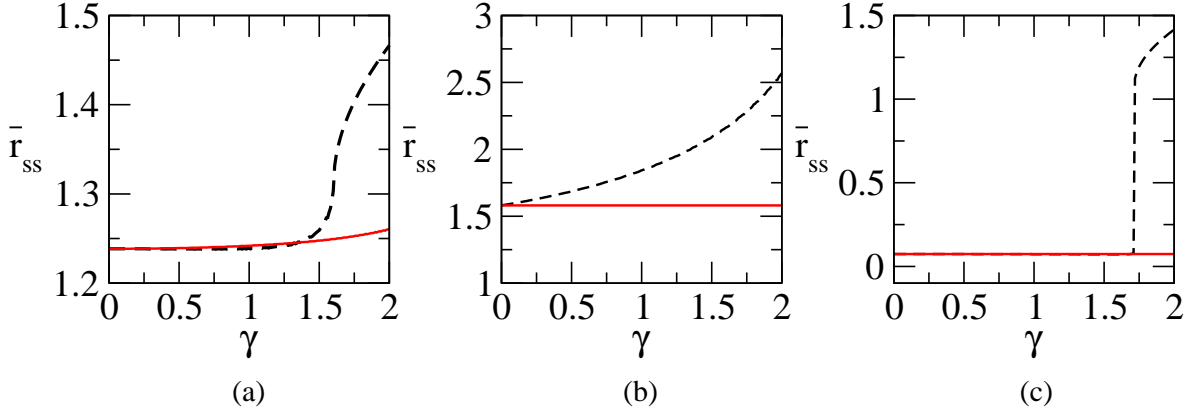}
\caption{Plots of scaled radius for different $\gamma$ with two different values of $\Omega=2$(black, doted) and $4$(red) where (a) is for the limit cycle situation having $a=1,b=0.2<\frac{\sin(\omega t_d)}{\omega}$, (b) is the center situation with $a=0,b=\frac{\sin(\omega t_d)}{\omega}$ and (c) is the center-type situation with $a=1,b=\frac{\sin(\omega t_d)}{\omega}$. The time delay $t_d, \epsilon$ and $\omega$ are fixed at 0.623, 0.05 and 1, respectively.} 	\label{fig ps}
\end{figure}

Figs. \ref{fig ps}(a), (b) and (c) describe the plots for amplitudes at steady state for the limit cycle, center and center-type cases, respectively, with $\gamma$ for two different values of $\Omega$ where the resonances are appearing. In fig.~\ref{fig ps}(a), for both $\Omega=2$ and $4$, amplitude increases with $\gamma$. For $\Omega=2$, amplitude saturates to a fixed value at steady state, but at $\Omega=4$, the amplitudes are oscillating at the steady state. The plot has been done by taking averages at  steady state for each $\gamma$. Similarly, fig.~\ref{fig ps}(b) shows, amplitude  increases very slowly with $\gamma$  for $\Omega=2$ and for $\Omega=4$, it is fixed and gives the center in nature. Likewise, fig.~\ref{fig ps}(c) shows, the amplitude increases with $\gamma$ and gives a limit cycle for $\Omega=2$, but for $\Omega=4$, the amplitude increases with a very small amount, showing center-type in nature. So, there is some switching transition at $\Omega=2$ in each case like center is converting into center-type orbits and center-type orbit is going to limit cycle state due to increase in $\gamma$.

\begin{figure}[!ht]
\centering
\includegraphics*[width=\linewidth]{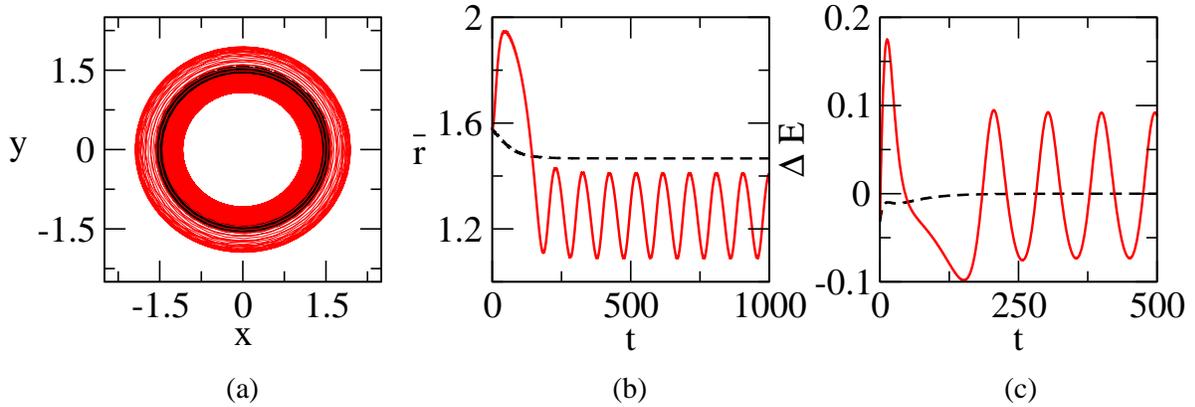}
\caption{\textbf{Limit Cycle Case:} Plots of (a) phase space, (b) scaled radius and (c) energy consumption per cycle for limit cycle due to parametric excitation for $\Omega=2$(black, doted) and $\Omega=4$(red). The fixed parameter values for this state are $a=1,~b=0.2<\frac{\sin(\omega t_d)}{\omega},~\omega=1,\gamma=2$ and $ \epsilon=0.05,$ where time delay $t_d$ is kept fixed in 0.623.}	\label{fig lc}
\end{figure}

Next, fig.~\ref{fig lc}(a) is the phase space plot of the limit cycle due to parametric resonances for $\Omega=2$ and $\Omega=4$. Fig.~\ref{fig lc}(b) describes the amplitudes of the limit cycle oscillators due to excitation, where $\Omega=2$ case settles down at steady value, but at $\Omega=4$, amplitude oscillates with two different periods and giving rise to  an oscillatory antiresonance. Plots of energy consumption per cycle for both the cases are given in fig.~\ref{fig lc}(c) which shows zero value at $\Omega=2$ as the amplitudes become saturated, but at $\Omega=4$, the curve oscillates as the amplitudes oscillate and average will be close to zero line. For van der Pol system the resonance and antiresonance behaviours are already studied in ref.~\cite{penvo}, but here we have obtained an oscillating antiresonance.

\begin{figure}[!ht]
\centering
\includegraphics*[width=\linewidth]{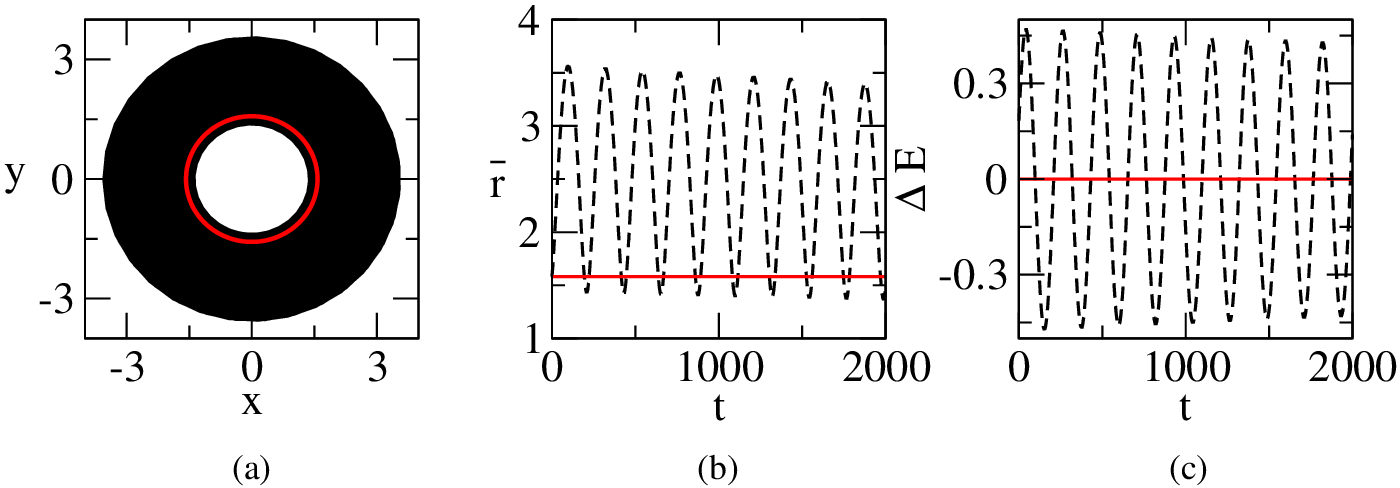}
\caption{\textbf{Center:} Plots of (a) phase space, (b) scaled radius and (c) energy consumption per cycle for limit cycle due to parametric excitation for $\Omega=2$(black, doted) and $\Omega=4$(red). The fixed parameter values for this state are $a=0,~b=\frac{\sin(\omega t_d)}{\omega},~\omega=1,\gamma=2,~t_d=0.623$ and $ \epsilon=0.05.$}	\label{fig c}
\end{figure}

Figs.~\ref{fig c}(a), (b) and (c) are the plots for phase space, amplitude variation with time and change in average energy of each cycle for the center while parametric excitation is acting upon it. Fig.~\ref{fig c}(b) shows there is no oscillation in amplitude at $\Omega=4$, but at $\Omega=2$ the amplitude oscillates and decaying with time and showing the periods are different in each cycle as a footprint. If we take the average of the amplitude in this case then it can be comparable  with the power law of decreasing amplitude and one can say that due to parametric excitation the center case is dramatically converted into center-type case  at the resonating mode, $\Omega=2$. Fig.~\ref{fig c}(c) shows that for both resonating modes it touches the zero line whereas at $\Omega=4$ it settles to a constant value from initial, but at $\Omega=2$ the dynamics  follows the power law behaviour.

\begin{figure}[!ht]
\centering
\includegraphics*[width=\linewidth]{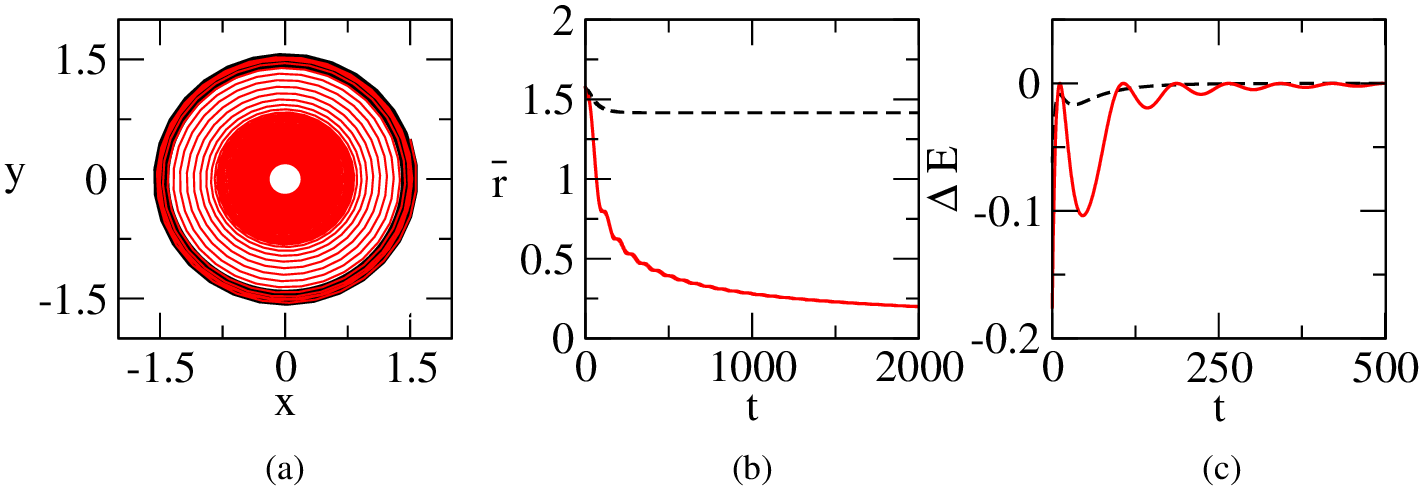}
\caption{\textbf{Center-type:} Plots of (a) phase space, (b) scaled radius and (c) energy consumption per cycle for limit cycle due to parametric excitation for $\Omega=2$(black, doted) and $\Omega=4$(red). The fixed parameter values for this state are $a=1,~b=\frac{\sin(\omega t_d)}{\omega},~\omega=1,\gamma=2,~t_d=0.623$ and $ \epsilon=0.05.$}	 \label{fig ct}
\end{figure}

Further, figs.~\ref{fig ct}(a), (b) and (c) are the plots for phase space, amplitude-time variation and energy consumption in each cycle for center-type case in presence of parametric excitation. The singularity at $\Omega=2$ gives a limit cycle oscillation which can be confirmed from fig.~\ref{fig ct}(b) where the amplitude touches a steady state value. But, at $\Omega=4$, it follows the power law decay in amplitude with time having a small amount of oscillation that can be fix by taking averages. Fig.~\ref{fig ct}(c) shows the plots for energy consumption per cycle in each modes where the variation touches the zero value within a small amount of time  for $\Omega=2$ and at $\Omega=4$, it shows quite oscillating in nature as amplitude oscillates.

\subsection{Stability Region of Parametrically Excited Limit Cycle}	\label{subsec2}
For weak time delay as in the earlier numerical and analytical scenario we have shown the stability region of the parametrically excited system. 
Relying on Eqs.~\eqref{eqo2} and \eqref{eqo4}, we can determine the steady state solution  and periodic solutions around it, in addition to their stability profile. Assuming that the coefficients $u$ and $v$ are small and vary as $u \approx e^{st}$ and $v \approx e^{st}$ one obtain relations for each of the system upon linearising the above Eqs.~\eqref{eqo2} and \eqref{eqo4}. At resonance, $\lbrace s-\frac{\epsilon}{2} \sin(t_d)\rbrace^2=\frac{\epsilon^2}{4}\left[b^2-\lbrace\cos(t_d)+\frac{b \gamma}{2}\rbrace^2\right]$  leads to $|\cos(t_d)+\frac{b \gamma}{2}|\le \cos(t_d)+\frac{|b \gamma|}{2}<|b|$, and at antiresonance, $\lbrace s-\frac{\epsilon}{2} \sin(t_d)\rbrace^2=\frac{\epsilon^2}{4}\lbrace b^2-\cos^2(t_d)\rbrace$  leads to $\cos(t_d)<|b|$.

The stability of the fixed point is determined by the eigenvalues of the Jacobian matrix of the vector fields in Eqs.~\eqref{eqo2} and \eqref{eqo4}. The characteristic polynomials which give the eigenvalues are given below(for resonance and antiresonance respectively)
\begin{align}
\lambda^2+ \epsilon \lambda (b-\sin \left(t_d\right))-\frac{1}{16} b^2 \gamma ^2 \epsilon ^2+\frac{b^2 \epsilon ^2}{4} -\frac{1}{2} b \epsilon ^2 \sin \left(t_d\right)+\frac{\epsilon ^2}{4}(\sin^2 (t_d)+\cos^2 (t_d))=0,	\label{eg2}
\end{align}
\begin{align}
\sigma^2+ \epsilon \sigma \left(b-\sin \left(t_d\right)\right)+\frac{b^2 \epsilon ^2}{4}-\frac{1}{2} b \epsilon ^2 \sin \left(t_d\right) +\frac{\epsilon ^2}{4}(\sin^2 (t_d)+\cos^2 (t_d))=0.	\label{eg4}
\end{align}
\normalsize
From Eqs.~\eqref{eg2} and \eqref{eg4}, it is clear that the Jacobian matrices of the vector fields at the initial equilibrium solutions both has two complex conjugate eigenvalues, namely $\lambda_{1,2}=A_1+iB_1$ and $\sigma_{1,2}=A_2+iB_2$ where $A_1=\epsilon (\sin(t_d)-b)/2=A_2=\alpha (say), B_1=\frac{\epsilon}{2} \sqrt{\cos^2(t_d)-\frac{b^2 \gamma^2}{4}}$ and $B_2=\frac{\epsilon \cos(t_d)}{2}$. Thereby, solving Eqs.~\eqref{eqo2} and \eqref{eqo4} near the origin is amounts to determine a dynamical phase-space trajectory in the form $(u, v) = e^{A_{1,2}t}(\cos B_{1,2}t, \sin B_{1,2}t)$. One can find  that the equilibrium state is locally stable for $\alpha < 0$, while it is unstable for $\alpha > 0$. For $\alpha=0$, the eigenvalues are imaginary with close orbital  path will be center or slowly decaying center-type situation. The zero critical value of the parameter, $\alpha$ determines a Hopf bifurcation point, where one encounters a significant qualitative change in the system's dynamical profile. In our case, one physically assumes $\alpha > 0$ with an unstable equilibrium state.

For unit frequency, the eigenvalues for resonance and antiresonance are $$\lambda_{1,2} \approx \pm 0.03125 \epsilon  \sqrt{64. b^2 \gamma ^2-168.847}-0.5 b \epsilon +0.291737 \epsilon$$  and  $$\sigma_{1,2} \approx -0.5 b \epsilon +(0.291737\, \pm 0.406066 i) \epsilon$$ (the eigenvalues of all $\Omega \neq 2$) respectively, where the values of $a$ and $t_d$ are $1$ and $0.623$ respectively. The respective eigenvalues give the nature of the above autonomous system as well as the original parametrically excited system. From the graphs, we find that $\Omega = 2$ is the resonant point and at the non-zero singular point $\Omega = 4$, there exist an oscillating antiresonance. Since, we have chosen $\omega=1$, the resonance came at $\Omega = 2$ as a parametric resonance appears at twice the eigen frequency.
\begin{figure}
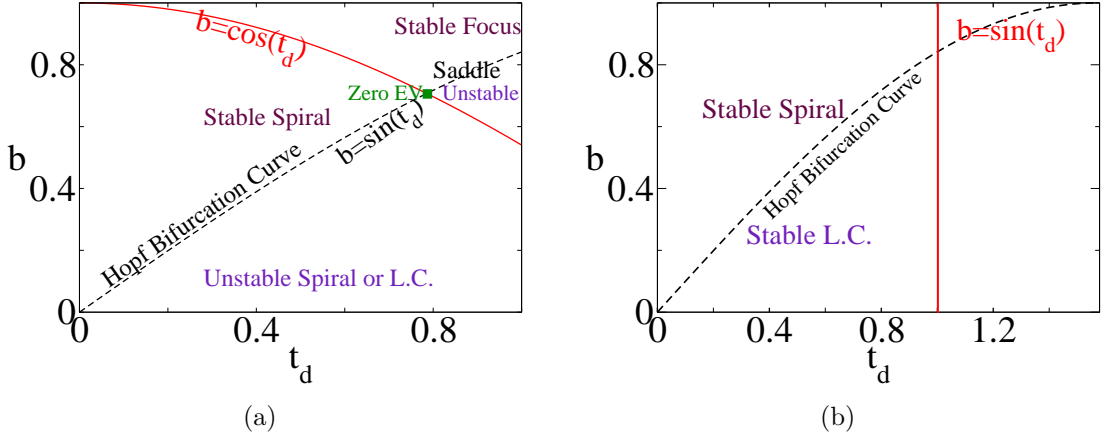

\centering
\subfigure[]{
\resizebox*{2.7in}{!}{\includegraphics*{f6a-EV_Space_CO_2.eps}} \label{fig7.1a}} 
\quad
\subfigure[]{
\resizebox*{2.7in}{!}{\includegraphics*{f6b-EV_Space_CO_4.eps}} \label{fig7.2b}}
\caption{Stability diagram for parametrically excited time delayed system for $b$ with time delay ($t_d$) is obtained from linear stability analysis for the cases (a) $\Omega=2$ and (b) $\Omega=4$.} \label{fig es}
\end{figure}
In fig.~\ref{fig es} stability diagram for parametrically excited time delayed system for $b$ with time delay ($t_d$) are shown from linear stability analysis for the limiting case (a) $\Omega=2$ and (b) $\Omega=4$.

\section{Stability Analysis and Hopf Bifurcation}  \label{sec4}
Here we have proposed stability analysis and Hopf bifurcation of the system for the full range of parameter space. We shall perform local stability analysis of the trivial fixed point, $E_0(0,0)$ and the existence of Hopf bifurcation of the model system \eqref{eq}. The linearized system at the origin is written as

\begin{align}
\left\{ \begin{array}{l}
\dot{x}(t) = y(t),	\\
\dot{y}(t) =  -\epsilon x(t - t_d) - \epsilon b \dot{x}(t) - {\omega^2} x(t).
\end{array} \right.  \label{15}
\end{align}
                                                                                                                                                                             
The corresponding characteristic equation of system \eqref{15} is given by
\begin{align*}
\lambda^2 + b \epsilon \lambda + \omega^2 + \epsilon e^{-\lambda t_d} = 0.
\end{align*} 
Above equation can be rewritten as
\begin{align}
(\lambda^2 + b \epsilon \lambda + \omega^2) e^{\lambda t_d} + \epsilon = 0.  \label{17}
\end{align} 
        
Now, we have the following cases:

\textbf{Case 1:} When $t_d =0$, Eq.~\eqref{17} becomes
\begin{align}
\lambda^2 + b \epsilon \lambda + \omega^2 + \epsilon = 0.  \label{18}
\end{align} 

We can see that the conditions for all the roots of Eq.~\eqref{18} are negative (as all the coefficients are positive) or to have negative real part is given by Routh-Hurwitz criterion as $b \epsilon>0$ and $\omega^2 +\epsilon>0$. Then, the trivial equilibrium point $E_0$ is locally asymptotically stable.

\textbf{Case 2:} When $t_d \neq 0$, $\lambda= i s$ be the root of Eq.~\eqref{17}, we have
\begin{align}
(-{s^2} + i s b \epsilon +\omega^2)(\cos(t_d s) + i \sin(t_d s)) +\epsilon = 0.	\label{17a}
\end{align}

Equating real and imaginary parts, we have
\begin{align}
(-s^2 + \omega^2) \cos(t_d s) -s b \epsilon \sin(t_d s)+ \epsilon =& 0, \notag \\
s b \epsilon \cos(t_d s) +(-s^2 + \omega^2) \sin(t_d s) =& 0. \notag
\end{align}
We have from above equation
\begin{align}
\cos(t_d s) =& \frac{\epsilon (s^2 -\omega^2)}{(-s^2 + \omega^2)^2 +(s b \epsilon)^2}=\frac{s^2-\omega^2}{\epsilon}, \notag \\
\sin(t_d s) =& \frac{s b \epsilon^2}{(-s^2 + \omega^2)^2 +(s b \epsilon)^2}=sb. 	\label{19}
\end{align}
Now, from the trigonometrical relation $\sin^2(t_d s) +\cos^2(t_d s) =1$, we have
\begin{align}
s^4 +(b^2 \epsilon^2 -2 \omega^2) s^2 + \omega^4 -\epsilon^2 = 0. \label{20}
\end{align}
Eq.~\eqref{20} has the following roots $s_j$, which is given by
\begin{align}
s_j = \pm \sqrt{ \frac{ 2 \omega^2 -b^2 \epsilon^2 \pm \epsilon \sqrt{b^2 (b^2 \epsilon^2 - 4 \omega^2) +4}}{2}}; \quad j=1,2,3,4.  \label{21}
\end{align}
The above Eq.~\eqref{21} has at least one positive root $s_0$ if the condition $ b < \frac{\sqrt{2}}{\epsilon} \sqrt{\omega^2 + \sqrt{\omega^4 -\epsilon^2}}$ satisfies.

We obtain the corresponding critical value of time delay $t_{d_l}$ for $s_0$ as
\begin{align}
t_{d_l} =  \left\lbrace \begin{array}{ll}
\frac{1}{s_0} \left( \sin^{-1} {b s_0} +2\pi l \right), & \text{if }  b < \frac{\sqrt{2}}{\epsilon} \sqrt{\omega^2 + \sqrt{\omega^4 -\epsilon^2}},  \\
\frac{1}{s_0} \left(\pi -\sin^{-1} {b s_0} +2\pi l \right), & \text{if }  b > \frac{\sqrt{2}}{\epsilon} \sqrt{\omega^2 - \sqrt{\omega^4 -\epsilon^2}},
\end{array} \right. \quad l= 0, 1, 2, \cdots.	\label{22}
\end{align}

Define $t_d^* = \min \lbrace t_{d_l} \rbrace$, i.e., $t_d^*$ is the smallest positive value of $t_{d_l}$; $l=0,1,2,\cdots$, given by the above Eq.~\eqref{22}. Now, we determine whether the roots of Eq.~\eqref{17} cross the imaginary axis of the complex plane as $t_d$ varies. Let $\lambda(t_d) = \xi(t_d) + i s(t_d)$ be a root of the Eq.~\eqref{17} such that these two conditions $\lambda(t_d) = \xi(t_{d_l}) =0$ and $s(t_{d_l})=s_0$ satisfies. 

\begin{lemma}
Transversality condition satisfies if the following holds
 $$\left[ \Re \left(\frac{d\lambda}{dt_d}\right)^{-1} \right]_{t_d =t_d^*} \neq 0.$$
\end{lemma}
\begin{proof}  
Differentiating Eq.~\eqref{17} with respect to $t_d$, we obtain
\begin{align}
\left( \frac{d\lambda}{dt_d} \right)^{-1} = \frac{b \varepsilon + i 2 s}{s (b \varepsilon s+ i (s^2-\omega^2))} +i \frac{t_d}{s}.	\notag
\end{align}
Now collecting the real parts at $t_d = t_d^*$, we have
\begin{align}
\left[ \Re \left( \frac{d\lambda}{dt_d} \right)^{-1} \right]_{t_d = t_d^*} = \frac{ (b \varepsilon)^2 +2 (s^2 - \omega^2) }{(b \varepsilon s)^2 + (s^2 - \omega^2)^2},	\label{22a}
\end{align}
which shows that 
\begin{align}
\left[ \Re \left( \frac{d\lambda}{dt_d} \right)^{-1} \right]_{t_d = t_d^*} >0, \text{ if } b < \frac{\sqrt{2}}{\epsilon} \sqrt{\omega^2 + \sqrt{\omega^4-\epsilon^2}}.  \notag
\end{align}
Therefore, the transversality condition is satisfied for each $t_d = t_d^*$ and hence Hopf bifurcation occurs at $t_d = t_d^*$. This completes the proof. 
\end{proof}
If $ \Re \left( \frac{d\lambda}{dt_d} \right)>0,$ then all those roots that crosses the imaginary axis with non-zero speed at $(\lambda=) i s$ from left to right as $t_d$ increases. Now, we state the following result:
\begin{theorem}
If trivial equilibrium point $E_0$ exists then that point of the model system \eqref{eq} is locally asymptotically stable when $t_d \in [0, t_d^*) $ and unstable for $t_d>t_d^* $. Furthermore, the system undergoes Hopf bifurcation at $E_0$ when $t_d = t_d^*,$ provided $b < \frac{\sqrt{2}}{\epsilon} \sqrt{\omega^2 + \sqrt{\omega^4-\epsilon^2}}$.
\end{theorem}

\textbf{Note:} For this section and the next section, we have considered $b$ analytically as a constant parameter only.

\subsection{Direction and Stability of Hopf Bifurcation} \label{subsec5} 
In this subsection, we will discuss the direction, stability and period of the bifurcating periodic solutions using normal form and center manifold theory, introduced by Hassard et al.~\cite{hassard}. We assume that the system \eqref{eq} about the fixed point $E_0$ undergoes Hopf bifurcation at the critical point $t_{d} =t_d^*$. Then $\pm is$ are corresponding purely imaginary roots of the characteristic equation at the critical point throughout this subsection. \\
Let $t_d =t_d^* +\mu$, where $\mu \in \mathbb{R}$. Define the space of continuous real valued functions as $\mathbb{C}= \mathbb{C}([-1,0], \mathbb{R}^2)$. Let $u_{1}(t)=x(t),~ u_{2}(t)=y(t)$ and ${u_i}(t)=u_{i}(t_d t)$ for $i=1,2$; the delay system \eqref{eq} then converts into functional differential equation in $\mathbb{C}$ as
\begin{align} 
\dot{u}(t) = L_{\mu} (u_t) + F(\mu, u_t),  \label{23}
\end{align} 
where $u(t)=(u_1(t)$, $u_2(t))^\top \in \mathbb{C}$, $u_t(\theta) = u(t+\theta) = (u_1(t+\theta)$, $u_2(t+\theta))^\top \in \mathbb{C}$ and $L_{\mu}: \mathbb{C} \to \mathbb{R}^2$, $F: \mathbb{R}\times \mathbb{C} \to \mathbb{R}^2$ are given respectively by 
\begin{align}
L_{\mu}(\phi)&=(t_d^* +\mu) \begin{pmatrix} 
0 & 1 \\ 
-\omega^2  & -b \epsilon
\end{pmatrix} \phi(0)+(t_d^*+\mu) \begin{pmatrix}
0 & 0 \\ 
-\varepsilon & 0
\end{pmatrix} \phi(-1), 	\label{24}  \\
F(\mu, \phi) &= (t_d^* +\mu) \begin{pmatrix}
0 \\  -a \epsilon \phi_1^2(0) \phi_2(0) 
\end{pmatrix},		\label{25}
\end{align} 
where $\phi = (\phi_1, \phi_2)^\top \in \mathbb{C}$. \\
By Riesz representation theorem, there exists a matrix $\eta(\theta,\mu)$, $\theta \in [0,1]$, whose components are of bounded variation functions such that 
\begin{align}
L_{\mu} \phi = \int_{-1}^0 d \eta(\theta,\mu) \phi(\theta), \quad \text{for } \phi \in \mathbb{C}.	\notag
\end{align}
By considering Eq.~\eqref{24}, one can choose 
\begin{align} 
\eta(\theta, \mu) = (t_d^* +\mu) \begin{pmatrix} 
0 & 1 \\ 
-\omega^2  & -b \epsilon
\end{pmatrix} \delta(\theta)+(t_d^*+\mu) \begin{pmatrix}
0 & 0 \\ 
-\epsilon & 0
\end{pmatrix} \delta(\theta+1),  \notag
\end{align} 
where $\delta$ is Dirac delta function.                                          
For $\phi \in \mathbb{C}$, define
\begin{align}
A(\mu)\phi(\theta) =\left\{\begin{array}{ll}\displaystyle{\frac{d \phi(\theta)}{d \theta}},& \theta \in [-1,0),\\
\\ \int_{-1}^0 d\eta(\theta,\mu) \phi(\theta), & \theta  = 0, \end{array}\right. \label{26} \\
R(\mu) \phi(\theta)  = \left\{\begin{array}{ll}
\begin{pmatrix}
 0 \\ 0 \end{pmatrix}, & \theta \in [-1,0), \\
F(\mu, \phi),& \theta=0.
\end{array}\right. 	\notag 
\end{align}
In order to convenient study of Hopf bifurcation problem, we transform system \eqref{23} into an operator equation of the form 
\begin{align}
\dot{u}(t) = A(\mu) u_t + R(\mu) u_t,	\label{27}
\end{align}
where $u_{t}(\theta)=u(t+\theta)$, $\theta \in [-1,0]$. \\
For $\psi \in \mathbb{C} ([0,1], (\mathbb{R}^2)^*)$, the adjoint operator $A^*$ of $A$ is defined by 
\begin{align}
A^*(\mu) \psi(m) = \left\{ \begin{array}{ll} -\frac{d\psi(m)}{dm}, & m\in(0,1], \\ \int_{-1}^0 \psi(-t) d\eta(t,0), & m=0. \end{array}\right.  \notag 
\end{align} 
For $\phi \in \mathbb{C} ([-1,0],\mathbb{R}^2)$ and $\psi \in \mathbb{C} ([0,1],(\mathbb{R}^2)^*)$, define the bilinear inner product in order to normalize the eigenvectors of operator $A$ and adjoint operator $A^*$. 
\begin{align}
\left\langle \psi(m), \phi(\theta) \right\rangle &= \bar{\psi}(0).\phi(0)-\int_{-1}^0 \int_{\xi = 0}^{\theta} \bar{\psi}^{\top} (\xi -\theta) d\eta(\theta) \phi(\xi) d\xi,	 \label{28} 
\end{align}
where $\eta(\theta)= \eta(\theta,0)$. Then $A$ and $A^*$ are adjoint operators. We know that $\pm i s_{0} t_d^*$ are eigenvalues of $A$. Therefore, they are also eigenvalues of $A^*$. Next we calculate the eigenvector $q(\theta)$ of $A(0)$ belonging to the eigenvalue $i s_{0} t_d^*$ and eigenvector $q^* (\theta)$ of  $A^*(0)$ belonging to the eigenvalue $-i s_{0} t_d^*$. \\
Then we have $ A(0) q(\theta) = i s_{0} t_d^* q(\theta)$  and $A^*(0) q^*(\theta) = -i s_{0} t_d^* q^*(\theta)$. Let $q(\theta) = (1,\alpha)^{\top} e^{i s_{0} t_d^* \theta}$ and $q^*(\theta) = P(1,\beta)^{\top} e^{-i s_{0} t_d^* \theta}$.
Thus, we can obtain
\begin{align} 
\alpha =i s_0, \quad \beta =\frac{i s_0}{\omega^2 +\epsilon e^{i s_{0} t_d^*}}.  \label{29} 
\end{align} 
From \eqref{28}, we have
\begin{align}
\langle q^*&(m),q(\theta) \rangle = \bar{q^*}(0).q(0)-\int_{-1}^{0} \int_{\xi=0}^{\theta} \bar{q^*}^{\top} (\xi-\theta) d\eta(\theta) q(\xi) d\xi,  \notag \\
&= \bar{P} (1+\alpha \bar{\beta}) -\int_{-1}^{0} \int_{\xi=0}^{\theta} \bar{P} \begin{pmatrix} 1 & \bar{\beta} \end{pmatrix}\times e^{-i t_d^* s_0 (\xi-\theta)} d\eta(\theta) \begin{pmatrix} 1 \\ \alpha \end{pmatrix} e^{i t_d^* s_0 \xi} d\xi, \notag \\
&= \bar{P} \bigg[ 1+\alpha \bar{\beta} \bigg. \left. -\int_{-1}^0 \begin{pmatrix} 1 & \bar{\beta} \end{pmatrix} \theta e^{i t_d^* s_0 \theta} \begin{pmatrix} 1 \\ \alpha \end{pmatrix} d\eta(\theta) \right],	\notag \\ 
&= \bar{P} \bigg[ 1+\alpha \bar{\beta} \bigg. \left. +t_d^* \begin{pmatrix} 1 & \bar{\beta} \end{pmatrix} \begin{pmatrix} 0 & 0 \\ -\epsilon & 0 \end{pmatrix}\begin{pmatrix} 1 \\ \alpha \end{pmatrix} e^{-i t_d^* s_0} \right],  \notag \\
&= \bar{P} [ 1+\alpha \bar{\beta} - \bar{\beta} \epsilon t_d^* e^{-i t_d^* s_0} ].
\end{align}
Thus, we can choose $\bar{P}$ as 
\begin{align} 
\bar{P} =& \frac{1}{ 1+\alpha \bar{\beta} -\bar{\beta} \epsilon t_d^* e^{-i t_d^* s_0}},  \label{30}
\end{align} 
then $ \langle q^*(m), q(\theta) \rangle = 1 $. Furthermore, $ \langle q^*(m), \bar{q}(\theta) \rangle = 0$. Now we obtain $q$ and $q^{*}$. \\
Next, we use the same notations as in~\cite{hassard} and we first compute the coordinates to describe the center manifold $C_{0}$ at $\mu=0$. Here, $u_{t}$ be the solution of \eqref{23} at $\mu =0$. Define
\begin{align}  
z(t)= \langle q^{*}, u_{t} \rangle, \label{31} 
\end{align} 
and then define 
\begin{align}
W(t,\theta) =& u_t (\theta) -z(t) q(\theta)-\bar{z}(t) \bar{q}(\theta), \notag \\
=&  u_t (\theta) -2 \Re \{ z(t) q(\theta) \}. \label{32} 
\end{align}
On center manifold $C_{0}$, we have $W(t, \theta) = W (z(t), \bar{z}(t),\theta)$, where                                                                                                            
\begin{align} 
W(z(t),\bar{z}(t),\theta) = W_{20} (\theta) \frac{z^2}{2} +W_{11} (\theta) z \bar{z} + W_{02} (\theta) \frac{\bar{z}^2}{2} + \cdots,  \label{33}
\end{align} 
$z$ and $\bar{z}$ are the local coordinates for center manifold $C_{0}$ in the direction of $q$ and $q^{*}$ respectively. Note that $W$ is real if $u_t$ is real. We consider only real solutions.\\
For the solution $u_t \in C_{0}$ of \eqref{23}, since $\mu =0$, we have
\begin{align}
\dot{z}(t) =& \langle q^*, \dot{u}_t \rangle,	\notag \\
=& \langle q^*, A(0) u_t +R(0)u_t \rangle,	\notag \\
=& i t_d^* s_0 z + \bar{q^*}(0).F(0, W(z,\bar{z},0) +2 \Re\{z(t)q(0)\}), \notag \\
\overset{\Delta}{=} & i t_d^* s_0 z+ \bar{q^*}(0).F_0(z,\bar{z}). \notag
\end{align}
Rewrite this equation as
\begin{align} 
\dot{z}(t) = i t_d^* s_0 z + g(z,\bar{z}), 	 \label{34}
\end{align} 
where $g(z,\bar{z}) = \bar{q^*}(0).F_{0}(z,\bar{z})$ and expand $g(z,\bar{z})$ in powers of $z$ and $\bar{z}$, that is 
\begin{align} 
g(z, \bar{z}) = g_{20} \frac{z^2}{2} + g_{11} z \bar{z} + g_{02} \frac{\bar{z}^2}{2} + g_{21} \frac{z^2 \bar{z}}{2}+ \cdots.   \label{35}
\end{align} 
We have
\begin{align}
g(z,\bar{z}) =& \bar{q^*}(0).F_{0}(z,\bar{z}),	\notag \\
=& \bar{P} t_d^* \begin{pmatrix} 1 & \bar{\beta} \end{pmatrix} \begin{pmatrix} 0 \\ -a \epsilon u_{1t}^2(0) u_{2t}(0) \end{pmatrix}, \notag 
\end{align} 
where $u_{t}(\theta) = (u_{1t}(\theta), u_{2t}(\theta))^\top = W(t,\theta)+z(t) q(\theta) +\bar{z}(t) \bar{q}(\theta)$ and $q(\theta) = (1,\alpha)^\top e^{i s_0 t_d^* \theta}$, then we have

\begin{align}
\begin{pmatrix} u_{1t}(\theta) \\ u_{2t}(\theta) \end{pmatrix} = \begin{pmatrix} W_{20}^{(1)}(\theta) \frac{z^2}{2} + W_{11}^{(1)}(\theta) z \bar{z}+W_{02}^{(1)}(\theta) \frac{\bar{z}^2}{2} + O(|z, \bar{z}|^3) \\ 
W_{20}^{(2)}(\theta) \frac{z^2}{2} + W_{11}^{(2)}(\theta) z \bar{z} + W_{02}^{(2)}(\theta) \frac{\bar{z}^2}{2} + O(|z,\bar{z}|^3) \end{pmatrix} + z \begin{pmatrix} 1 \\ 
\alpha \end{pmatrix} e^{i t_d^* s_0 \theta} + \bar{z} \begin{pmatrix} 1 \\ 
\bar{\alpha} \end{pmatrix} e^{-i t_d^* s_0 \theta},  \notag
\end{align} 
For $\theta=0$, we have
\begin{align}
u_{1t}(0) =& z + \bar{z} + W_{20}^{(1)}(0) \frac{z^2}{2} + W_{11}^{(1)}(0) z \bar{z} + W_{02}^{(1)}(0) \frac{\bar{z}^2}{2} +O(|z,\bar{z}|^3),  \notag \\
u_{2t}(0) =& \alpha z +\bar{\alpha} \bar{z} +W_{20}^{(2)}(0) \frac{z^2}{2} +W_{11}^{(2)}(0) z \bar{z}  +W_{02}^{(2)}(0) \frac{\bar{z}^2}{2} +O(|z,\bar{z}|^3).  \notag 
\end{align}
It follows that
\begin{align}
g(z,\bar{z}) =& \bar{P} t_d^* \begin{pmatrix} 1, & \bar{\beta} \end{pmatrix} \begin{pmatrix} 0, & -a \epsilon \left(z + \bar{z} + W_{20}^{(1)}(0)\frac{z^2}{2} +\cdots \right)^2 \left(\alpha z + \bar{\alpha} \bar{z} + W_{20}^{(2)}(0) \frac{z^2}{2} +\cdots \right) \end{pmatrix}^\top, \notag \\
=& - a \epsilon \bar{\beta} \bar{P} t_d^* (2\alpha +\bar{\alpha}) z^{2} \bar{z} + \cdots. \notag
\end{align}

Comparing the coefficients with \eqref{35}, we have
\begin{align}
g_{20} =& g_{11} = g_{02} = 0, \notag \\  
g_{21} =& -2 a \epsilon \bar{\beta} \bar{P} t_d^* (2 \alpha + \bar{\alpha}).  \label{36}
\end{align} 
Thus, we can compute the following quantities:
\begin{align}
C_1(0) =& \frac{i}{2s_0 t_d^*} \left(g_{20} g_{11} - 2 |g_{11}|^2 - \frac{|g_{02}|^2}{3} \right) + \frac{g_{21}}{2},  \notag \\
\mu_2 =& -\frac{ \Re \{C_1(0) \}}{\Re\{ \lambda'_{0}(t_d^*) \}},  \notag \\
\beta_2 =& 2 \Re \{C_1(0)\},  \notag \\
T_2 =& -\frac{ \Im \{C_1(0) \} + \mu_2 \Im \{ \lambda'_0(t_d^*) \}}{s_0 t_d^*}. \label{37} 
\end{align}
Above formulae give a description of Hopf bifurcation periodic solutions of \eqref{23} at $t_d = t_d^*$ on the center manifold. Notations $\mu_2$, $\beta_2$ and $T_2$ determine respectively the direction of Hopf bifurcation, stability and period of bifurcating periodic solutions~\cite{hassard}. We summarize the following theorem.
\begin{theorem} \label{4.3}
For expressions given in \eqref{37}, following results hold
\begin{enumerate}
\item[(i)] If $\mu_2 > 0, \, (\mu_2<0)$, then Hopf bifurcation is supercritical, (subcritical) and the bifurcating periodic solutions exist for $t_d > t_d^*, \, (t_d<t_d^*)$.
\item[(ii)] The bifurcating periodic solutions are stable if $\beta_2 < 0$ and unstable if $\beta_2 > 0$.
\item[(iii)] The period of the bifurcating periodic solutions increases if $T_2 > 0$ and decreases if $T_2 < 0$.
\end{enumerate}
 \end{theorem}

\begin{example}
For the following model
\begin{align} 
\dot{x}(t) =& y(t), \notag \\ 
\dot{y}(t) =& -0.05 (x(t-0.623)+(x^{2}(t) +0.58347)y(t))-x(t). \notag
\end{align} 
 By taking $a=1, \omega =1, \epsilon =0.05, b=0.583474$ for the model system \eqref{7}. 
We have from Eq.~\eqref{37},
\begin{align*}
\mu_2 =&  0.7563576 > 0,  \\
\beta_2 =& -0.0308725 < 0, \\
T_2 =& 0.0174092 >0.
\end{align*} 
Therefore, from Theorem \ref{4.3}, we conclude that Hopf bifurcation is supercritical and bifurcating periodic solution is stable with increasing period.
\end{example}

\section{Bifurcation Analysis: Exact Numerical Simulation} 	\label{sec5}
\begin{figure}[!ht]
\centering
\subfigure[]{
\resizebox*{2.2in}{!}{\includegraphics{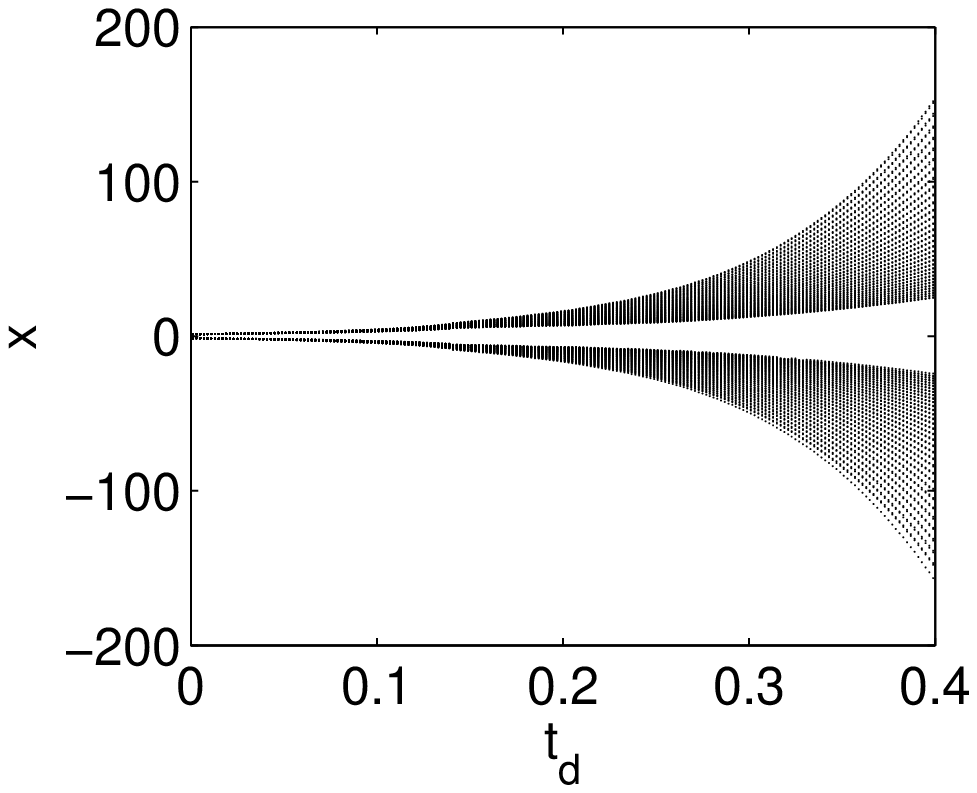}} \label{fig8.1a}} 
\quad
\subfigure[]{
\resizebox*{2.2in}{!}{\includegraphics{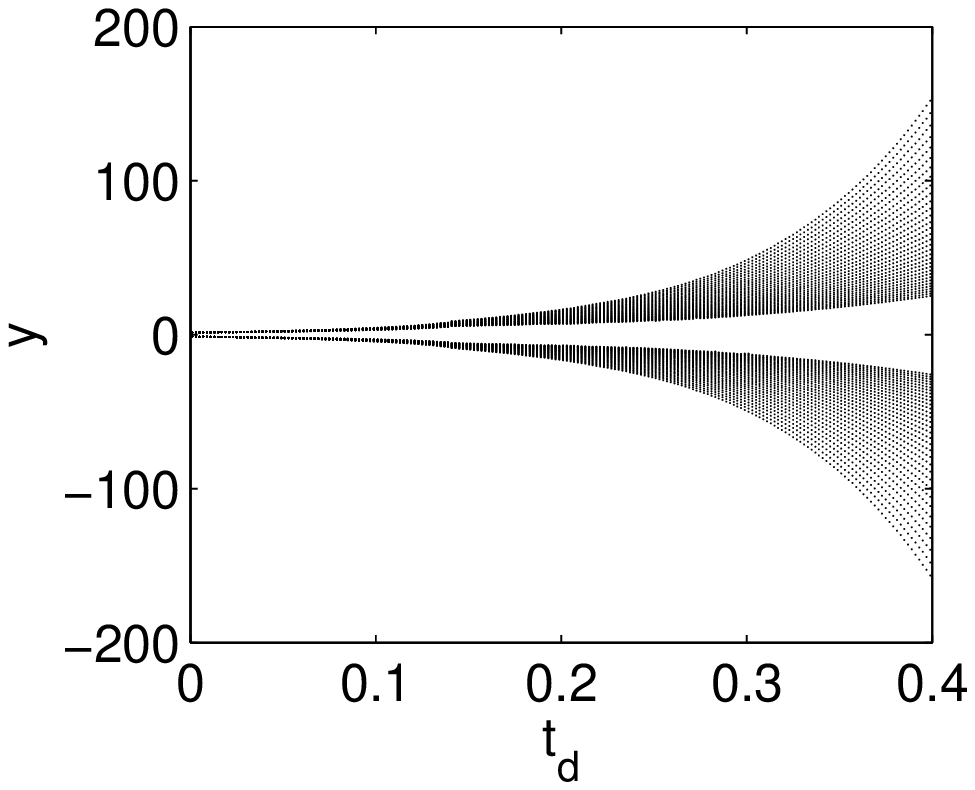}} \label{fig8.2b}} 
\caption{Bifurcation diagram for the model system \eqref{7} with the parameter values $\epsilon =0.05, \omega=1$, $a = 0, b = 0$. \textbf{(Feedback system with increasing phase space area)}} \label{fig8}
\end{figure}

\begin{figure}
\centering
\subfigure[]{
\resizebox*{1.35in}{!}{\includegraphics{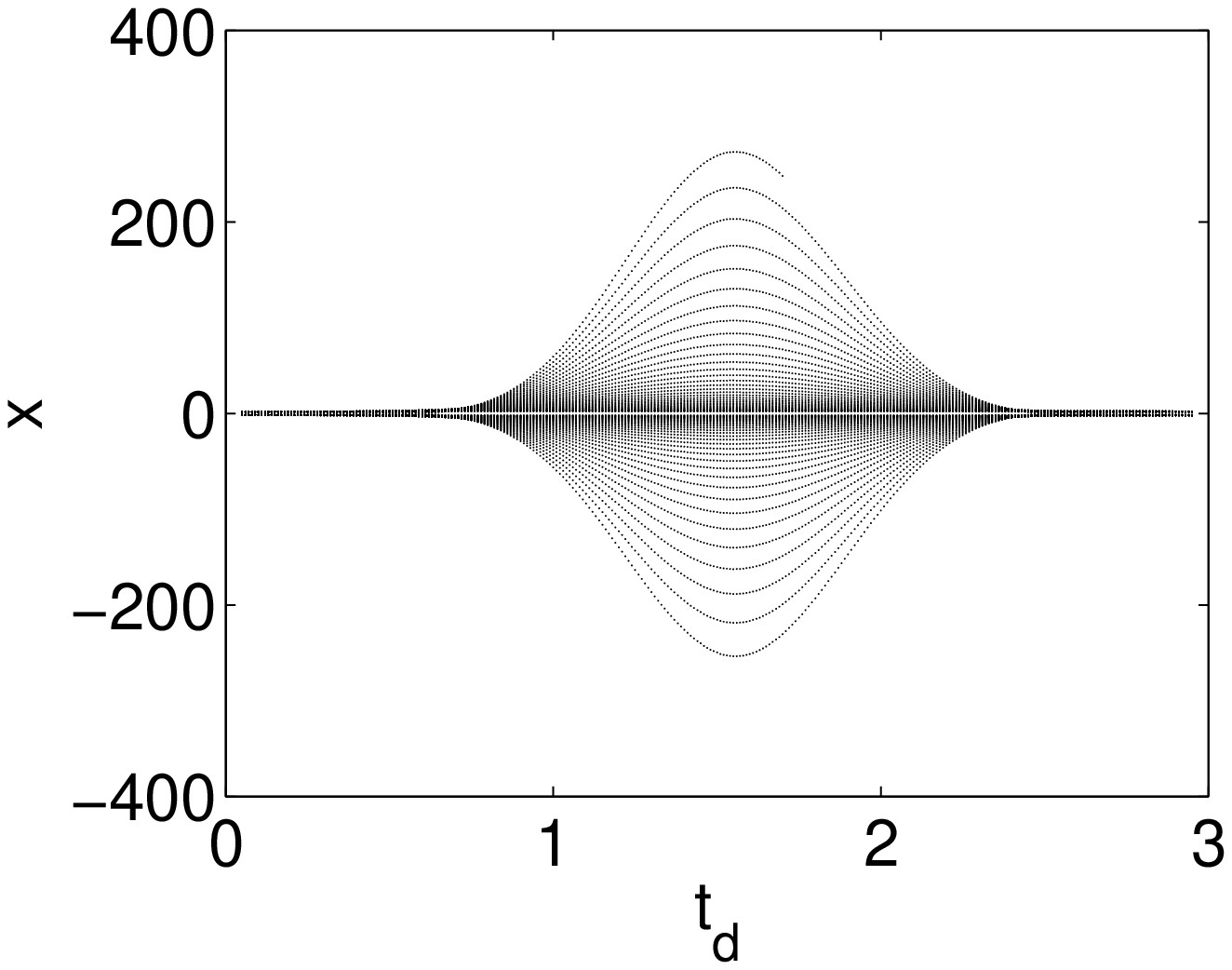}} \label{fig9.1a}} 
\quad
\subfigure[]{
\resizebox*{1.35in}{!}{\includegraphics{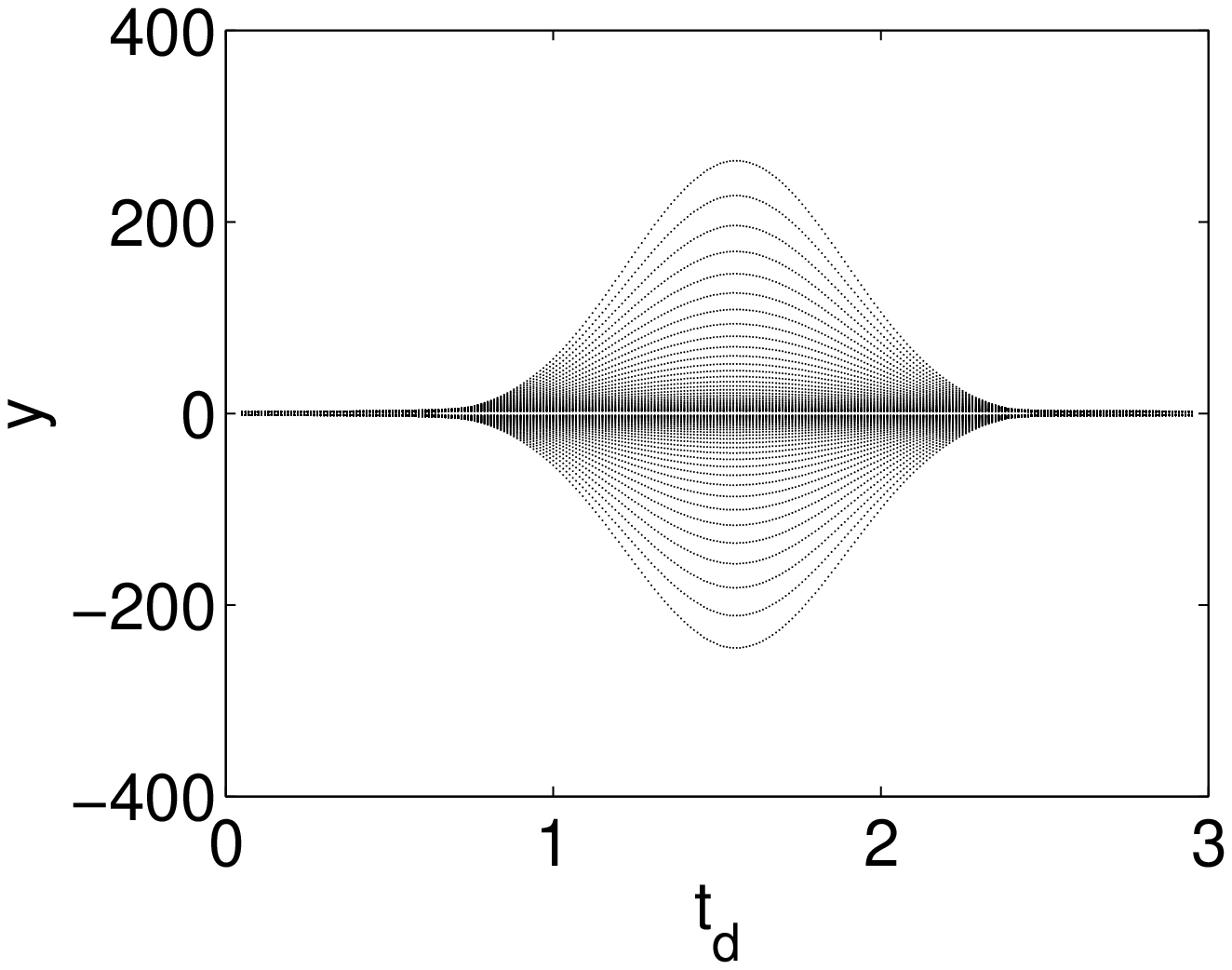}} \label{fig9.2b}} 
\quad
\subfigure[]{
\resizebox*{1.35in}{!}{\includegraphics{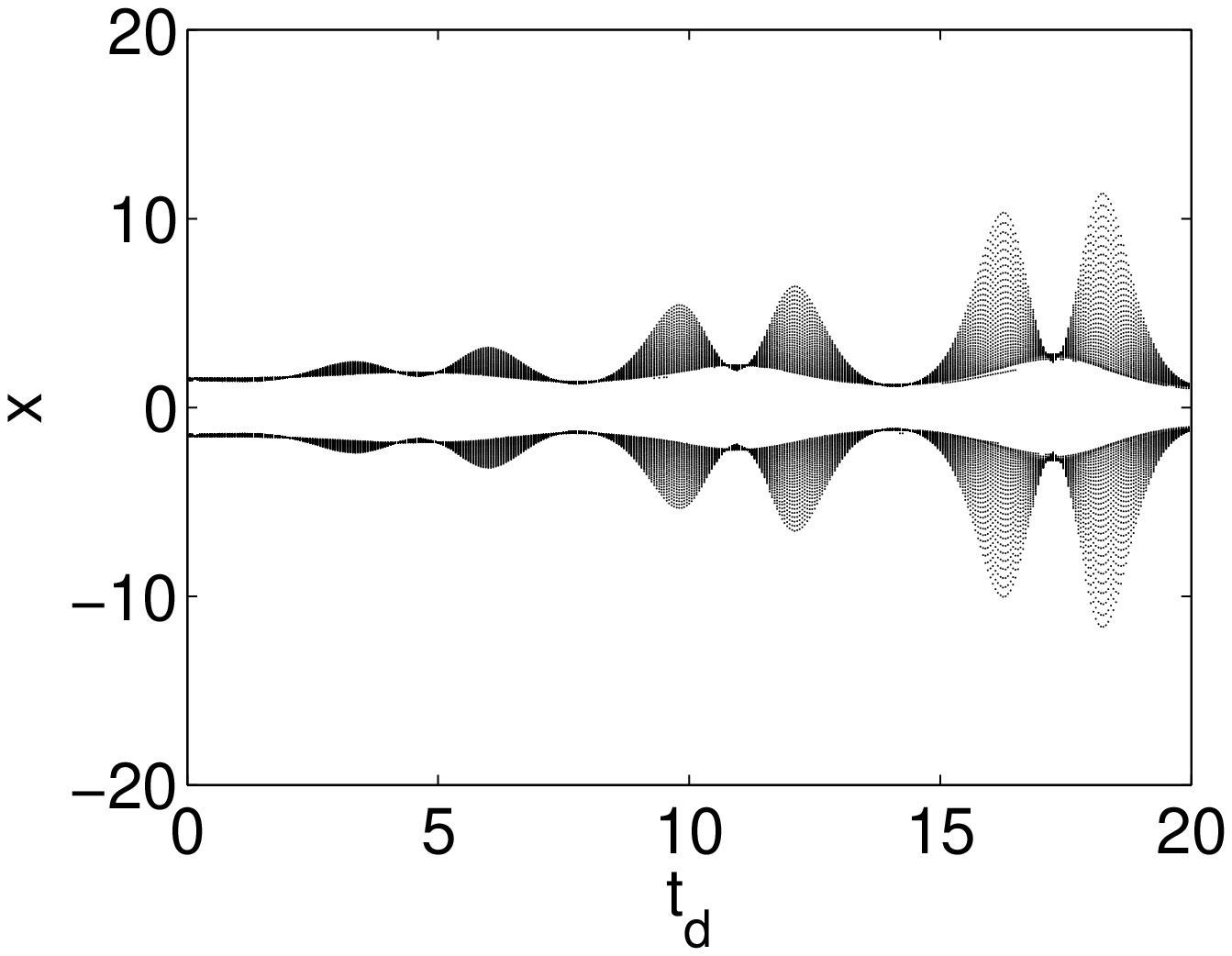}} \label{fig9.3c}} 
\quad
\subfigure[]{
\resizebox*{1.35in}{!}{\includegraphics{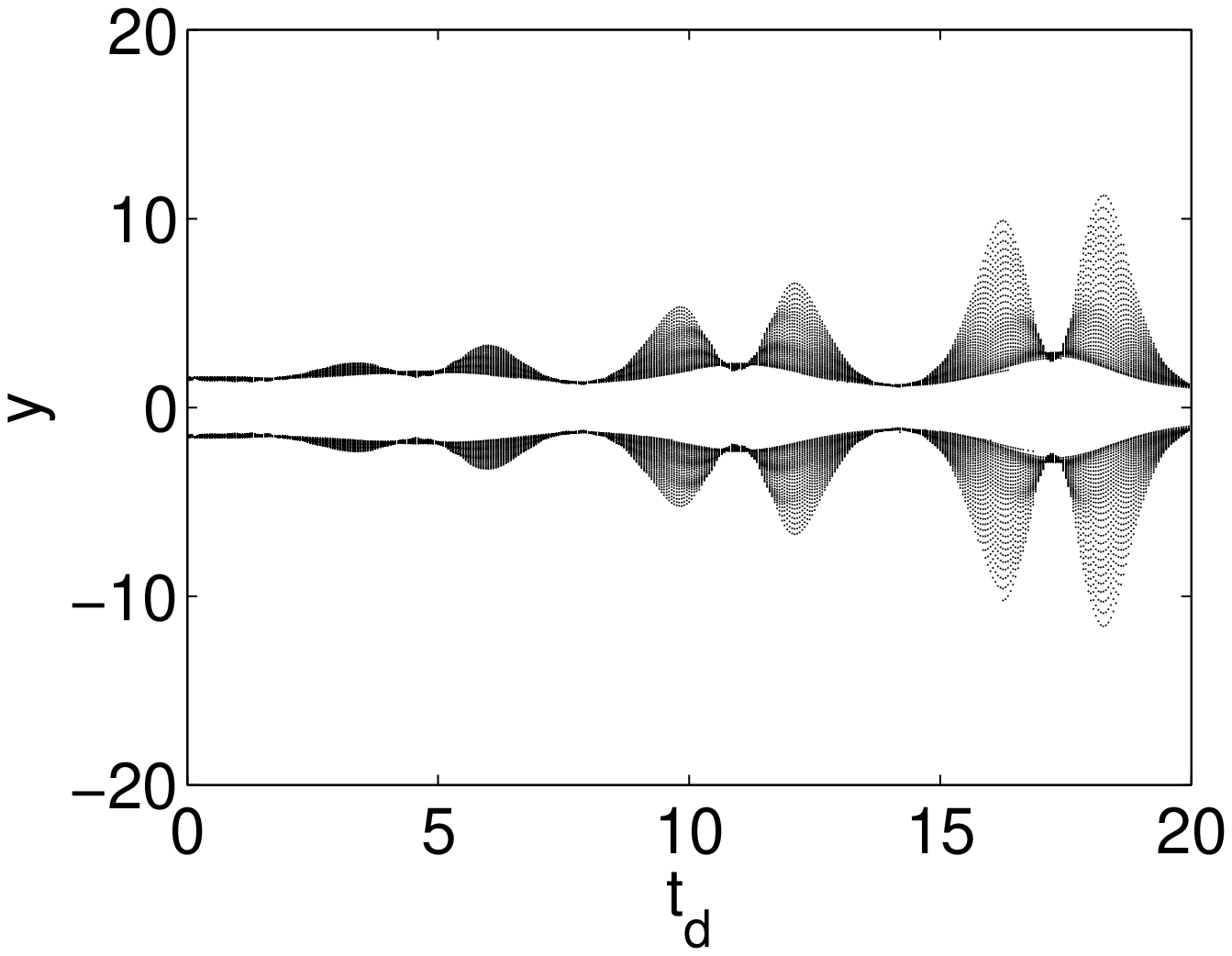}} \label{fig9.4d}} 
\caption{Bifurcation diagram for the model system \eqref{7} with the parameter values $\epsilon =0.05, \gamma = 2, \omega=1$, $a=0, b = \frac{\sin(\omega t_d)}{\omega}$. For figures (a) and (b) $\Omega=2$, and (c) and (d) $\Omega=4$. \textbf{(Center)}} \label{fig9}
\end{figure}

\begin{figure}
\centering
\subfigure[]{
\resizebox*{1.35in}{!}{\includegraphics{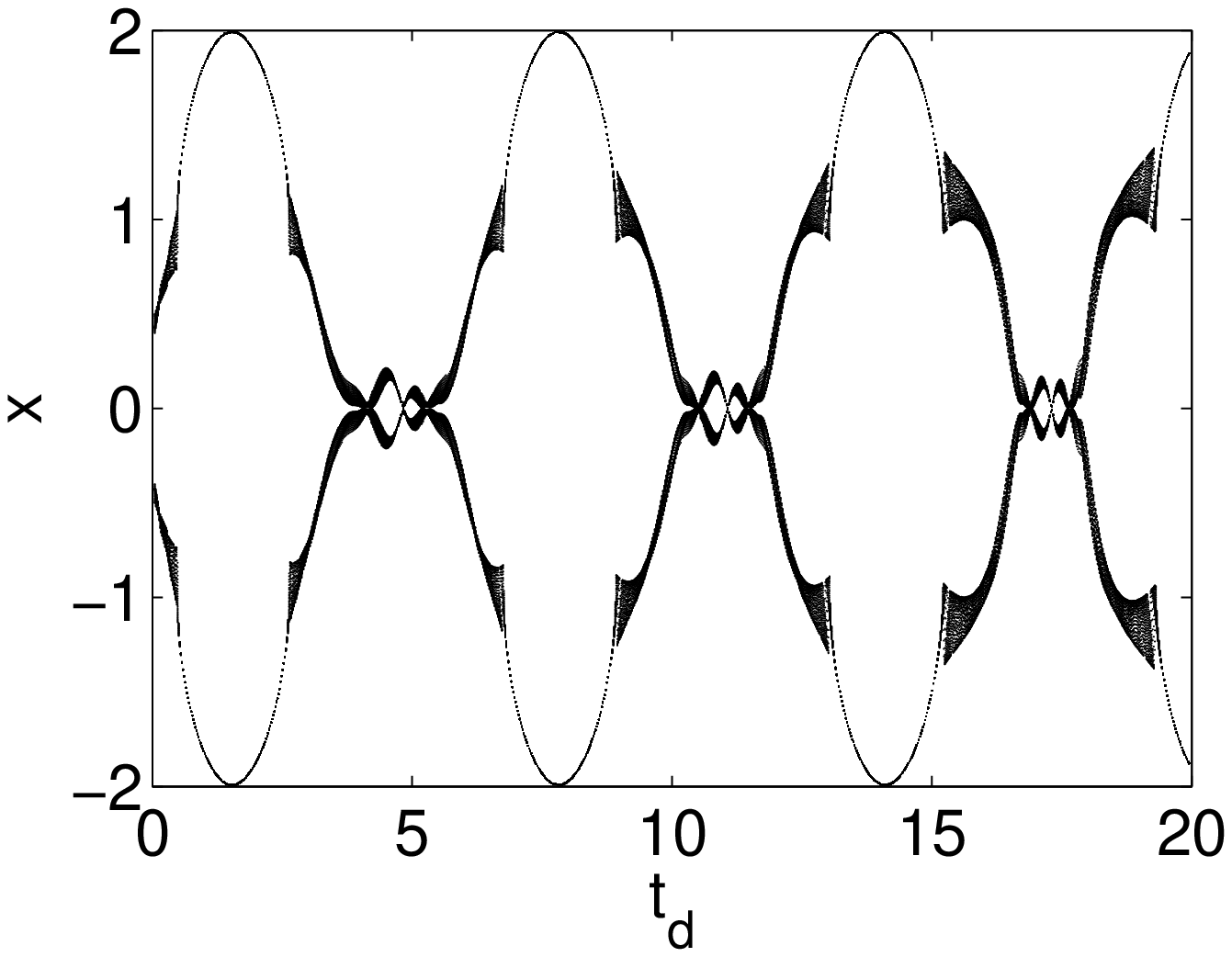}} \label{fig10.1a}} 
\quad
\subfigure[]{
\resizebox*{1.35in}{!}{\includegraphics{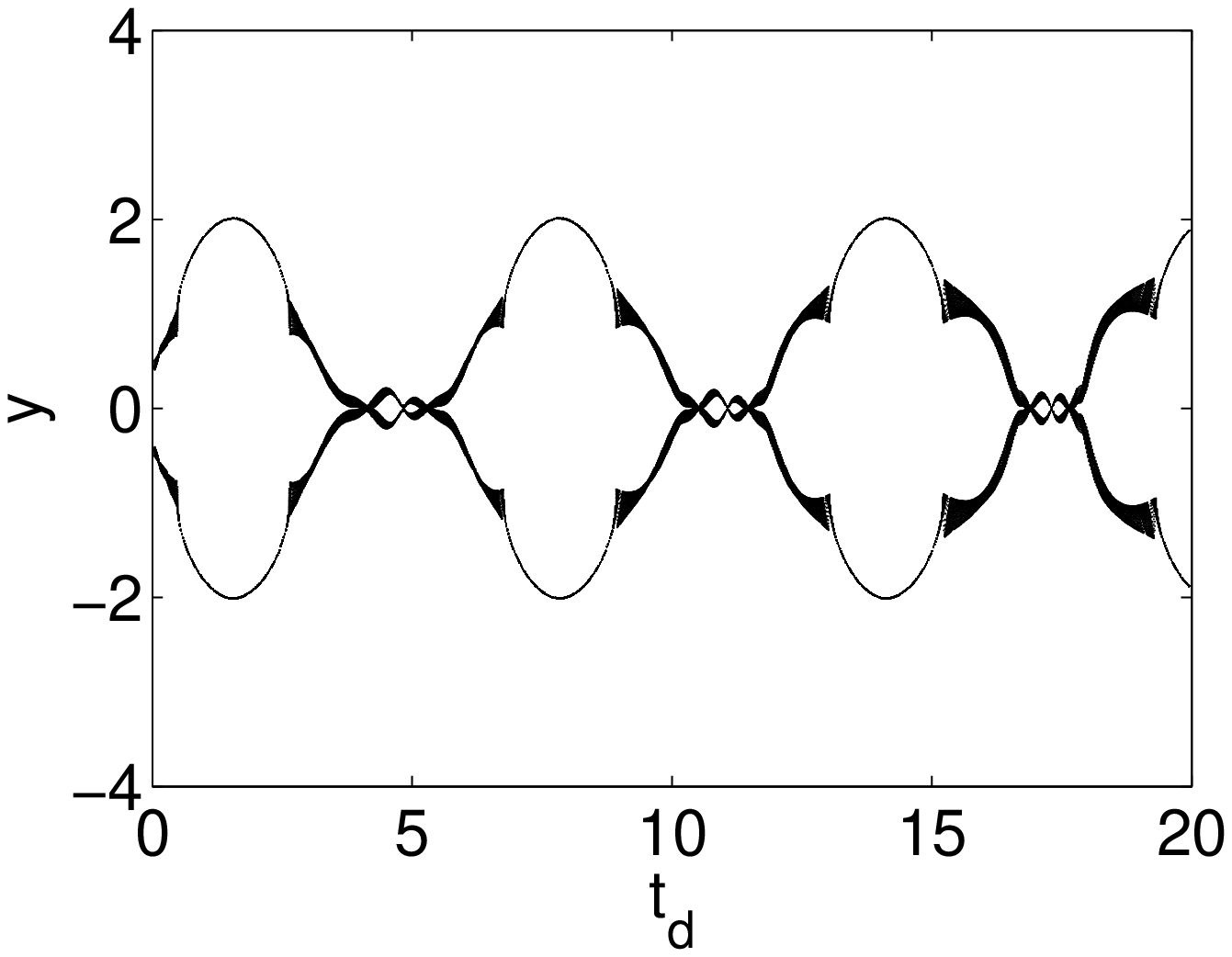}} \label{fig10.2b}} 
\quad
\subfigure[]{
\resizebox*{1.35in}{!}{\includegraphics{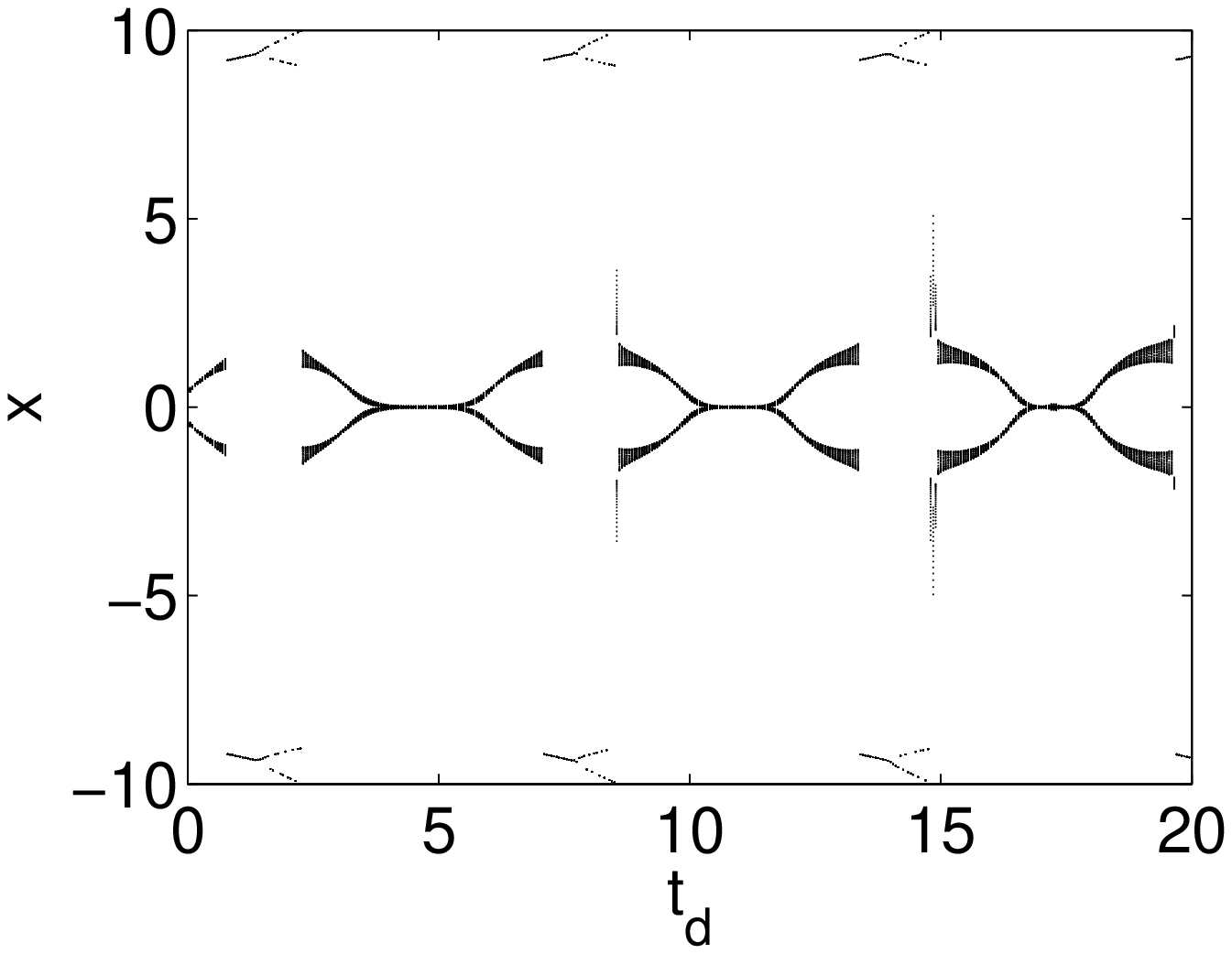}} \label{fig10.3c}} 
\quad
\subfigure[]{
\resizebox*{1.35in}{!}{\includegraphics{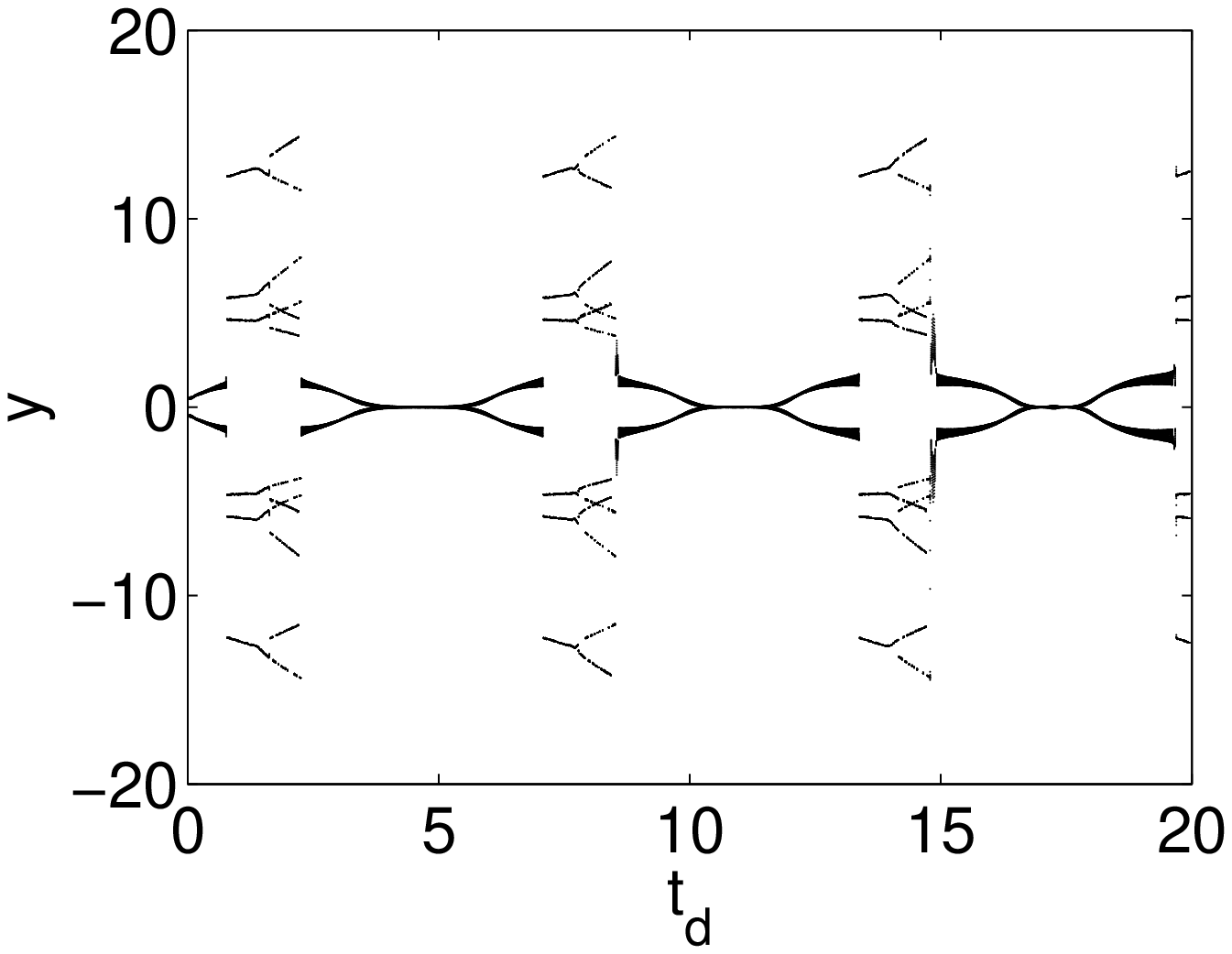}} \label{fig10.4d}}  
\caption{Bifurcation diagram for the model system \eqref{7} with the parameter values $\epsilon =0.05, \gamma = 2, b= \frac{\sin(\omega t_d)}{2 \omega}, \omega=1$, $a=1, b < \frac{\sin(\omega t_d)}{\omega}$. For figures (a) and (b) $\Omega=2$ and (c) and (d) $\Omega=4$. \textbf{(Limit cycle)}} \label{fig10}
\end{figure}

\begin{figure}
\centering
\subfigure[]{
\resizebox*{1.35in}{!}{\includegraphics{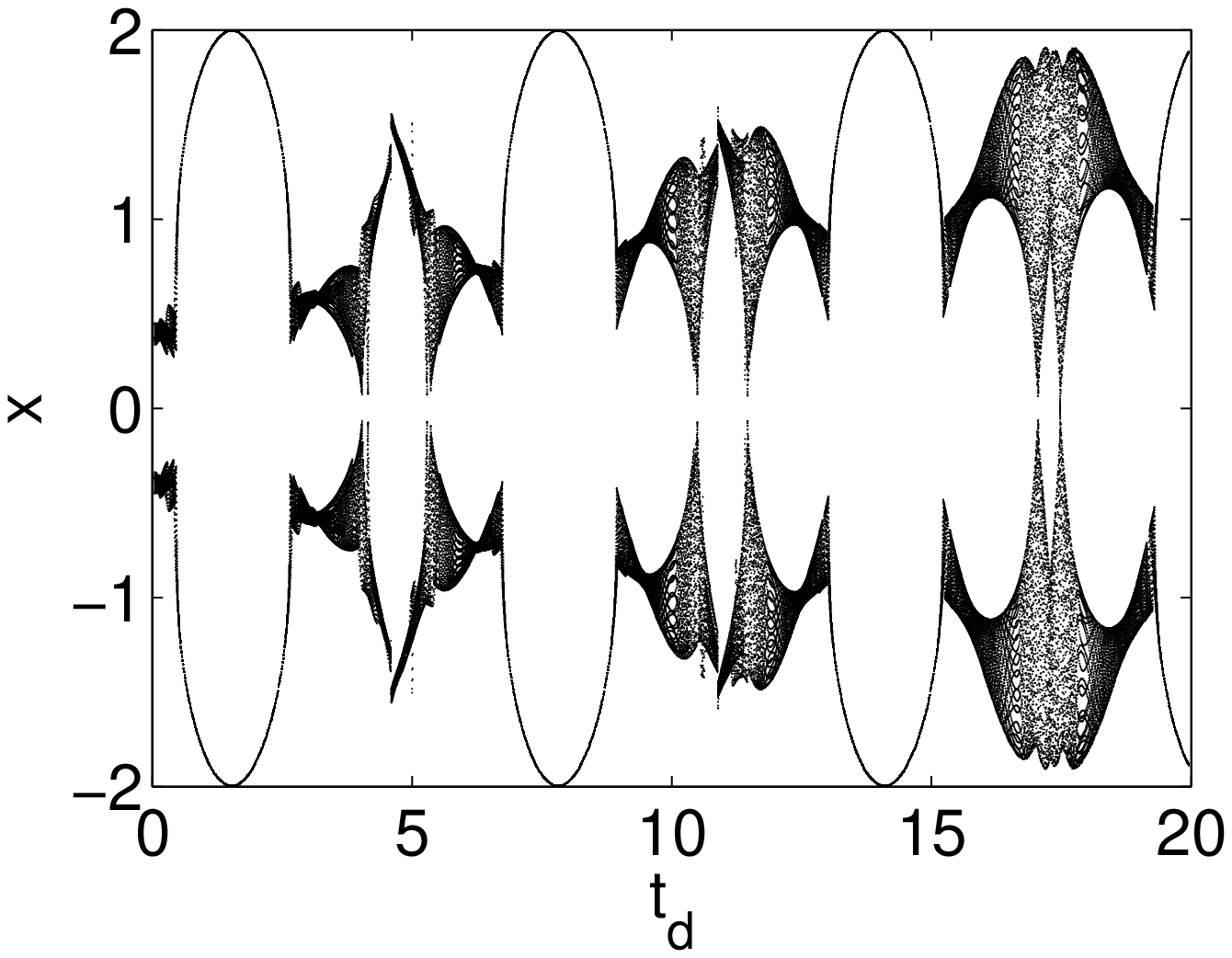}} \label{fig11.1a}} 
\quad
\subfigure[]{
\resizebox*{1.35in}{!}{\includegraphics{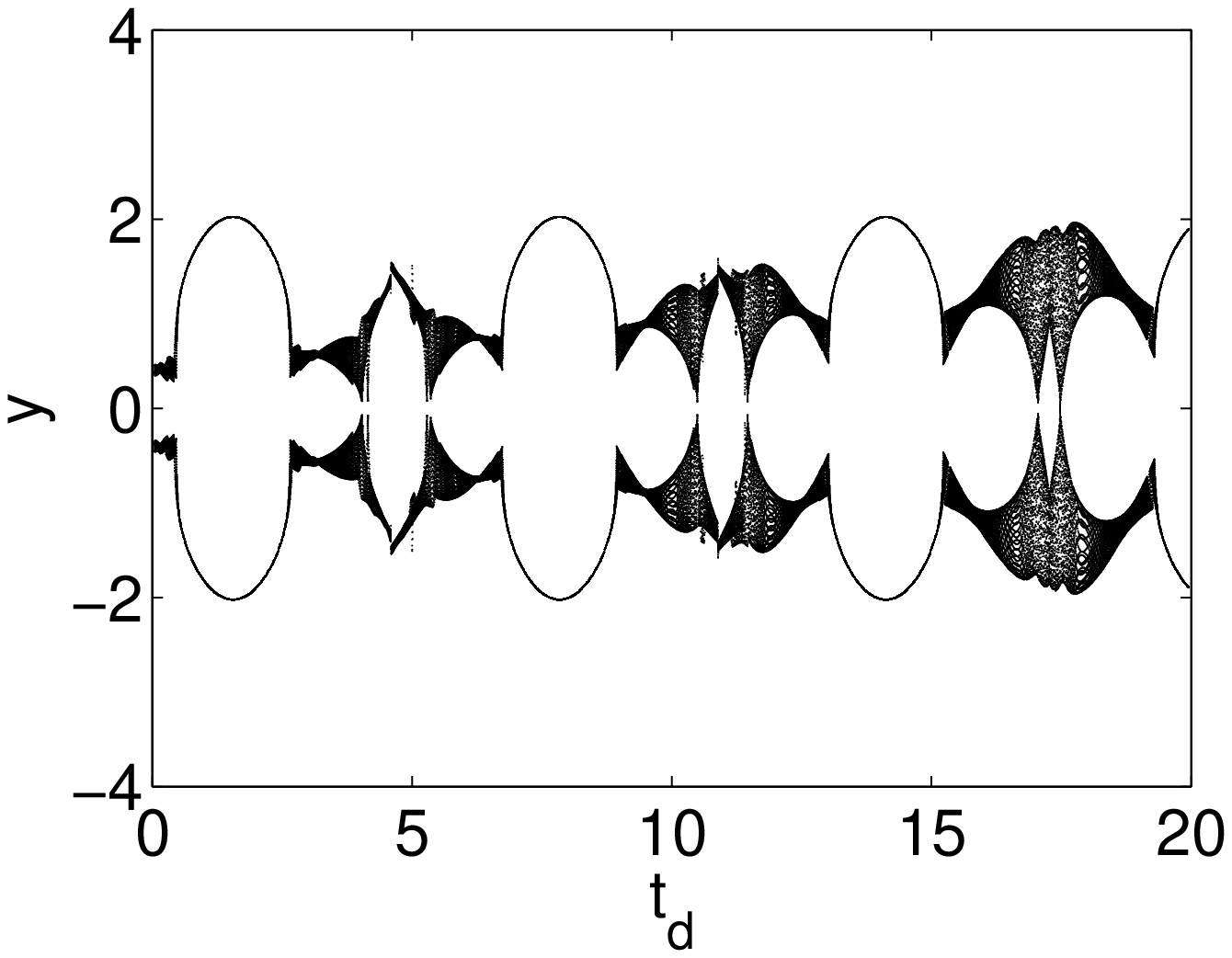}} \label{fig11.2b}}  
\quad
\subfigure[]{
\resizebox*{1.35in}{!}{\includegraphics{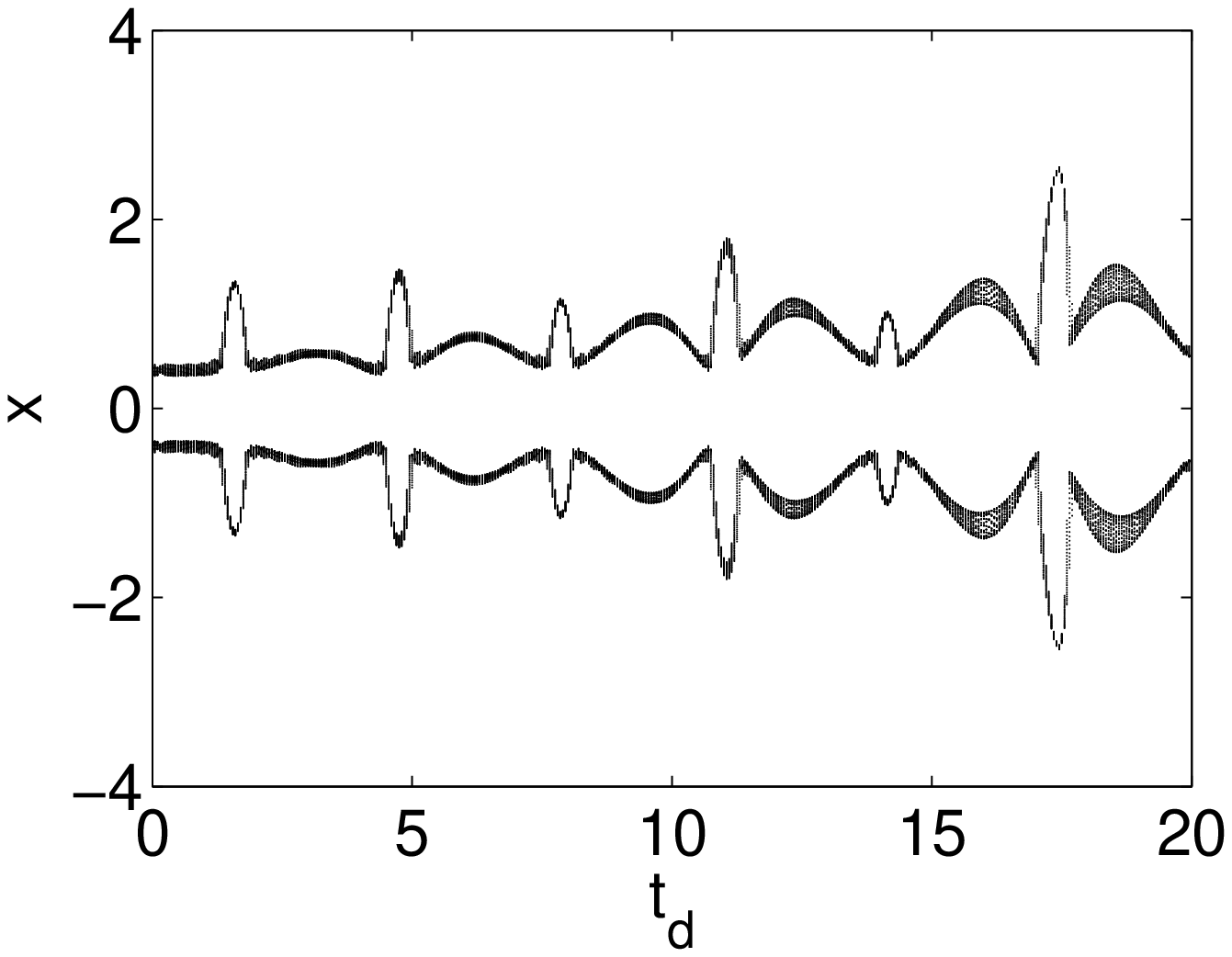}} \label{fig11.3c}} 
\quad
\subfigure[]{
\resizebox*{1.35in}{!}{\includegraphics{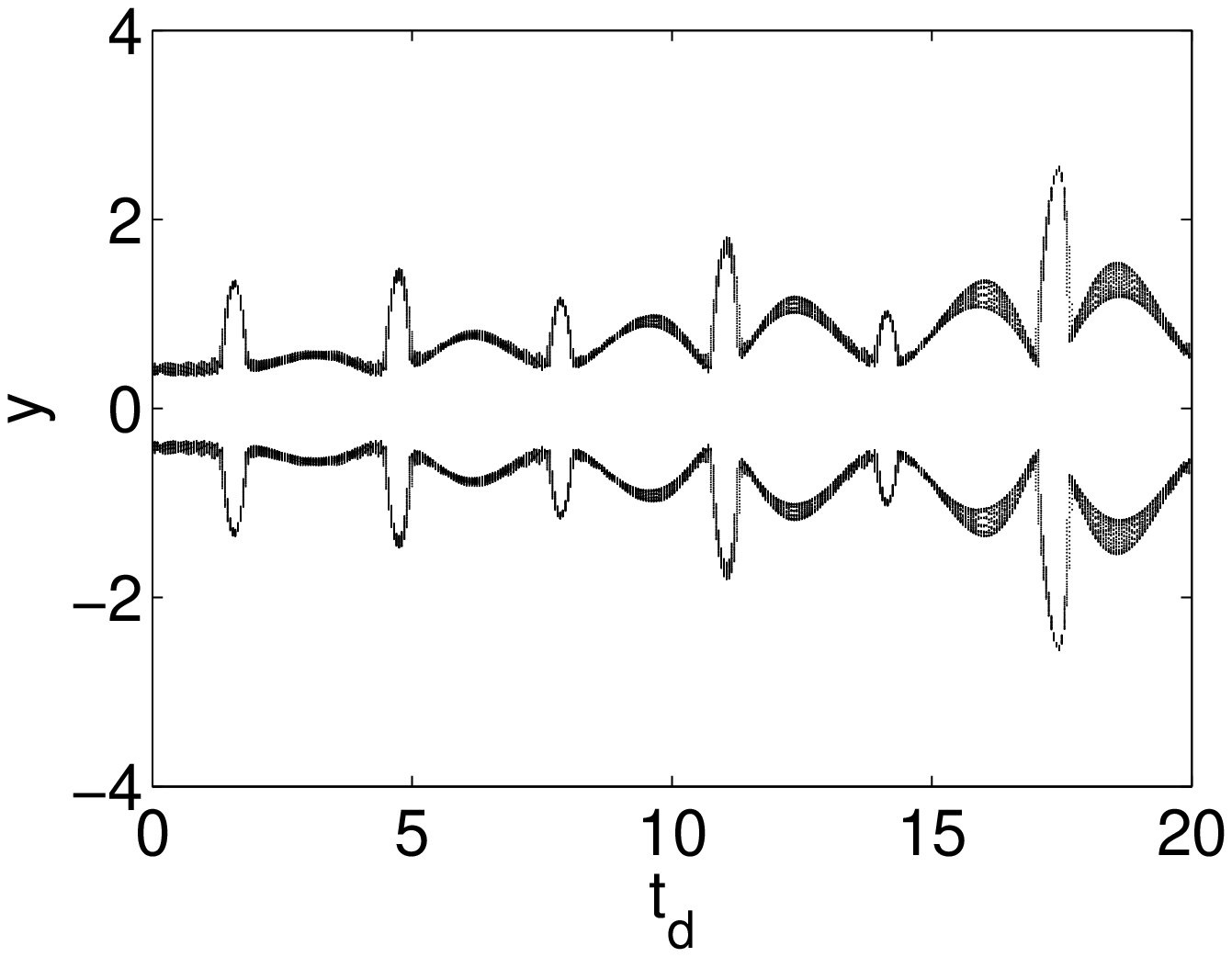}} \label{fig11.4d}}
\caption{Bifurcation diagram for the model system \eqref{7} with the parameter values $\epsilon =0.05, \gamma = 2, \omega=1$, $a=1, b = \frac{\sin(\omega t_d)}{\omega}$. For figures (a) and (b) $\Omega=2$, and (c) and (d) $\Omega=4$. \textbf{(Center-type orbit)}} \label{fig11}
\end{figure}

To investigate the effects of nonlinearity ($a$) and damping term ($b$), we carried out detailed bifurcation analysis of the model system \eqref{7}. Our main objective is to detect the existence of complex system dynamics in the presence of nonlinear damping. The system \eqref{7} is integrated using Matlab software for different cases of nonlinearity and damping with resonance ($\Omega=2$) and antiresonance ($\Omega=4$). For analysing the exact range of stability in details, we have represented bifurcation plots for both the state variables throughout the simulation. System dynamics show symmetric property throughout the simulation for both resonance and antiresonance cases. Time span is $[0,220]$ for fig.~\ref{fig9} and $[0,500]$ for figs.~\ref{fig8}, \ref{fig10} and \ref{fig11}.


\begin{figure}
\centering
\subfigure[]{
\resizebox*{1.35in}{!}{\includegraphics{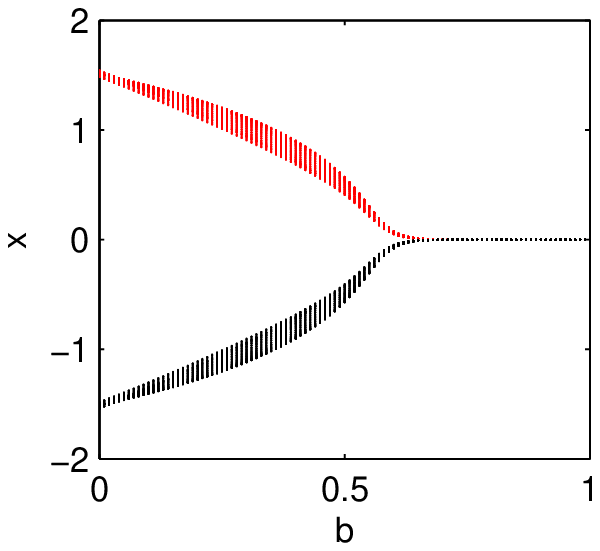}} \label{fig12.1a}} 
\quad
\subfigure[]{
\resizebox*{1.35in}{!}{\includegraphics{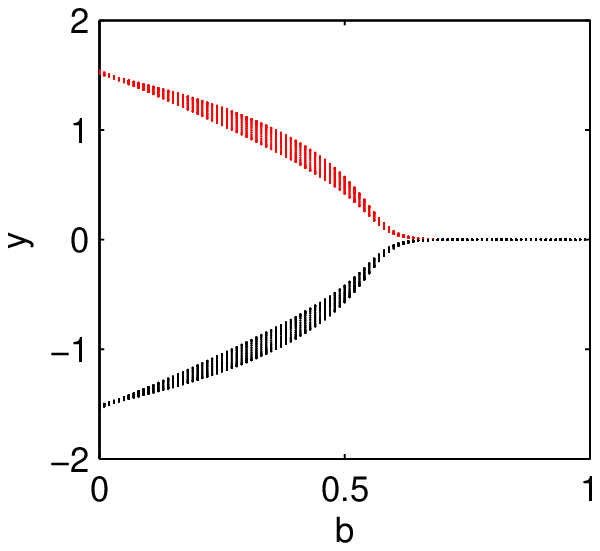}} \label{fig12.2b}}  
\quad
\subfigure[]{
\resizebox*{1.35in}{!}{\includegraphics{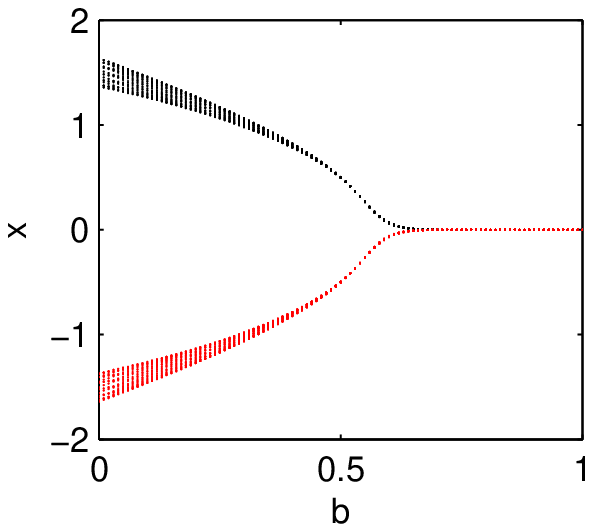}} \label{fig12.3c}} 
\quad
\subfigure[]{
\resizebox*{1.35in}{!}{\includegraphics{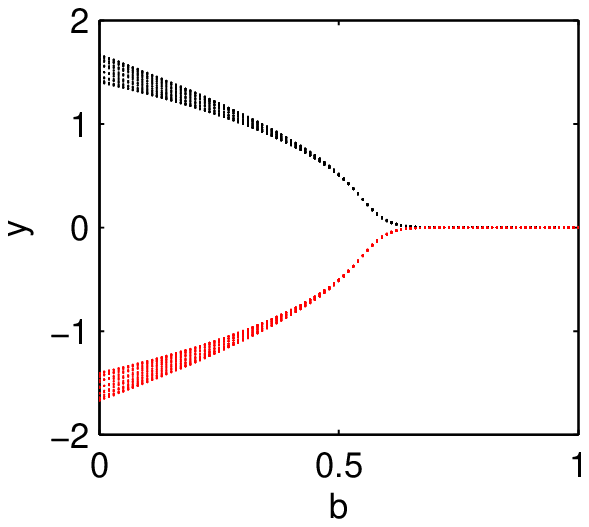}} \label{fig12.4d}} 
\caption{Bifurcation diagram of $b$ with the parameter values $\epsilon =0.05, \gamma = 1, a=1, \omega=1, t_d=0.623$. For figures (a) and (b) $\Omega=2$, and (c) and (d) $\Omega=4$.} \label{fig12}
\end{figure}

In fig.~\ref{fig8}, first we consider the case when $a=0$, $b=0$ i.e., no nonlinearity and damping terms present in the system. In this case, bifurcation depends only on the feedback controller $\epsilon$ as the frequency $\omega$ is equal to $1$. Feedback system lies in the range $[-200,200]$, can be observed from fig.~\ref{fig8}. Initially system is stable and as delay increases system becomes unstable.

We have investigated the effect of damping term $b = \frac{\sin(\omega t_d)}{\omega}$ in the absence of nonlinearity for the resonance case in figs.~\ref{fig9.1a}-\ref{fig9.2b} and antiresonance in figs.~\ref{fig9.3c}-\ref{fig9.4d}. System dynamics exhibits gap dependent bifurcation for both cases. The state variables lie in the domain $[-400,400]$ and $[-20,20]$, and $t_d \in (0,3]$ and $t_d \in (0,20]$ for resonance and antiresonance cases respectively. Phase space area of resonance case is larger than the antiresonance case. We observe center solution exists for the system \eqref{7} as it is highly dependent on damping terms.

In fig.~\ref{fig10}, we consider the effect of both nonlinear term  $a=1$ and damping term  $b = \frac{\sin(\omega t_d)}{2 \omega}$. Bifurcation diagram for resonance and antiresonance cases are executed in figs.~\ref{fig10.1a} and \ref{fig10.2b} respectively. System variables lie between $[-4,4]$ and $[-20,20]$, and $t_d \in (0,20]$ and shows number of stability switching scenario in fig.~\ref{fig10}. Sequences of period doubling of order 2 and 4 with symmetricity can be remarked for resonance in figs. \ref{fig10.1a}-\ref{fig10.2b} whereas for antiresonance, repeated scenario of period doubling and inverse period doubling of order 2, 6 and 12 are perceived in figs. \ref{fig10.3c}-\ref{fig10.4d}. Rich period doubling and halving scenario confirms the presence of limit cycles. We observe oscillatory dynamics which converges to a steady state for the system \eqref{7} from the bifurcation diagrams. 

In fig.~\ref{fig11}, nonlinearity and damping terms are $a=1$ and $b = \frac{\sin(\omega t_d)}{\omega}$ respectively. Resonance case is considered in figs.~\ref{fig11.1a}-\ref{fig11.2b} and antiresonance in figs.~\ref{fig11.3c}-\ref{fig11.4d}. System variables lie between $[-4,4]$ and $t_d \in [0,10]$  and shows center-type solution in bifurcation plots for both the cases. As $t_d$ increases, $x$ and $y$ also increases for non-resonance case. However, in resonance case $x$ and $y$ does not increases with $t_d$, it attains a maximum saturated peak and repeat it. Sequences of stability switches take places; period-doubling and halving layouts appear or disappear. As $t_d$ increases, the point of projection of period-doubling and halving occurs, exhibits more dense plot; such as it tending towards the chaotic scenario.

In fig.~\ref{fig12}, initially system shows limit cycle behaviour with $a=1, ~\gamma=1$ and in the range of $b \in [0,0.583]$ and stable for $b>0.583$. Hopf bifurcation occurs for both the resonance and non-resonance cases shown in figs.~\ref{fig12.1a}-\ref{fig12.2b} and \ref{fig12.3c}-\ref{fig12.4d}, respectively. Hence, we observed that the dynamics of the system bifurcates at the $b=0.583$.

\section{Discussions and Conclusions}		 \label{sec7}

A delay model in a damped quartic nonlinear oscillator is solved by multiscale perturbation method to obtain various periodic orbits, namely a limit cycle, center and a slowly decaying center with reference to a van der Pol oscillator having a limit cycle. This delay induced periodicity and bifurcation are probed through the parametric resonance and antiresonance. {\color{black}The calculation of response function due to a parametric excitation of an arbitrary periodic orbit is carried out here through K-B approach which is much handier than RG method specially in the context of  results obtained for the direction of Hopf bifurcation and stability of the bifurcating periodic solutions.} The effect of control parameters such as damping and nonlinear terms are investigated  via bifurcation analysis using normal form and center manifold theory.
\begin{enumerate}

\item We have found the characteristics of resonances due to parametric excitation. The nature of the resonances at $\Omega=2\omega$ and $\Omega=4\omega$ are investigated for limit cycle, center and center-type cases with $\omega=1$.

\item Stability criteria of the parametrically excited system for $\Omega=2 \omega$ and $4\omega$ resonances are investigated  with $\omega=1$ and they correspond to the approximate solution using K-B method.

\item  Linear stability analysis and bifurcation scenario for the full range of parameter space of the system \eqref{eq} is accomplished for the trivial fixed point. Occurrence of Hopf bifurcation of the fixed point ${E_0}(0,0)$ at critical point $t_d^*=0.623$  has been shown at which trivial point looses its stability. The values $\mu_2 = 0.7563576, \beta_2 = -0.0308725$ and $T_2 = 0.0174092$ indicates that Hopf bifurcation is supercritical and the bifurcating periodic solutions are stable with increasing periods.  

\item Stability and direction of Hopf bifurcation have been investigated using center manifold and normal form theory. We have also concluded from the bifurcation diagram, the system is unstable initially only for limit cycle solutions and stable for all the other solutions at the trivial point.

\item When the periodic orbit is a limit cycle in our system it can be clearly identified for the weakly delayed van der Pol case where the sign of $b$ and coefficient of $x(t-t_d)$ are both $<0$. 
\end{enumerate}

Thus, we can find that the damping is one of most effective parameter which stabilizes the system and delay plays a crucial role to lead the system unstable. In the presence of both damping and nonlinearity with delay stabilizes the system. The possibility of stabilizing a feedback delayed system with parametric excitation have a great effect in control of periodic flows, stabilization of high-speed milling in material formation and cutting process via spindle speed variation~\cite{stepan}.

\section*{Acknowledgements}
Sandip Saha acknowledges RGNF, UGC, India for the partial financial support. This work is supported by the Council of Scientific and Industrial Research (CSIR), Govt. of India under grant no. 25(0277)/17/EMR-II to R.K. Upadhyay. SS and GG are thankful to Dr. Sagar Chakraborty for useful comments.

\section*{References}
%

\bibliographystyle{unsrt}       
\bibliography{References}{}

\end{document}